\documentclass[onecolumn]{IEEEtran}

\usepackage[T1]{fontenc}
\usepackage[utf8]{inputenc}
\usepackage{graphicx}
\usepackage{amsmath,amssymb}
\usepackage{multirow} 
\usepackage{url}
\usepackage{float}
\usepackage{xcolor} 
\usepackage{booktabs}
\usepackage[acronym, shortcuts]{glossaries}
\makeglossaries
\newacronym{DBMC}{DBMC}{diffusion-based molecular communication}
\newacronym{TDMA}{TDMA}{time-division multiple access}
\newacronym{MDMA}{MDMA}{molecule-division multiple access}
\newacronym{NOMA}{NOMA}{non-orthogonal multiple access}
\newacronym{ADMA}{ADMA}{amplitude-division multiple access}
\newacronym{MA}{MA}{multiple access}
\newacronym{BEP}{BEP}{bit error probability}
\newacronym{TX}{TX}{transmitter}
\newacronym{RX}{RX}{receiver}
\newacronym{ISI}{ISI}{inter-symbol interference}
\newacronym{MAI}{MAI}{multiple-access interference}
\newacronym{SIC}{SIC}{successive interference cancellation}
\newacronym{OOK}{OOK}{on-off-keying}
\newacronym{MCS}{MCS}{Monte Carlo simulation}
\newacronym{SNR}{SNR}{signaling-molecule-to-noise ratio}
\newacronym{MC}{MC}{molecular communication}
\newacronym{IoBNT}{IoBNT}{Internet of Bio-Nano-Things}
\newacronym{BNM}{BNM}{bio-nano-machine}
\newacronym{DBMC-NOMA}{DBMC-NOMA}{NOMA for DBMC networks}
\newacronym{UCA}{UCA}{uniform concentration assumption}
\newacronym{CIR}{CIR}{channel impulse response}
\newacronym{SIMO}{SIMO}{single-molecule}
\newacronym{MUMO}{MUMO}{multi-molecule}
\newacronym{bps}{bps}{bits-per-second}
\newacronym{BER}{BER}{bit error rate}
\newacronym{ZF}{ZF}{zero-forcing}
\newacronym{MIMO}{MIMO}{multiple-input multiple-output}
\glsdisablehyper

\title{A Hands-On Molecular Communication Testbed for Undergraduate Education}

\author{
    \IEEEauthorblockN{Arne Gaedeken$^1$, Alexander Wietfeld$^2$, Yaning Zhao$^1$, Christian Deppe$^1$, Eduard Jorswieck$^1$, Wolfgang Kellerer$^2$\\}
    \IEEEauthorblockA{
        $^1$ Institute for Communications Technology (IfN)\\
        Technische Universität Braunschweig, Germany\\
        $^2$ Chair of Communication Networks \\
        Technical University Munich, Germany
    }
}

\begin{document}

\maketitle

\begin{abstract}
This work presents a hands-on molecular communication (MC) testbed developed for the undergraduate \textit{Communication Engineering} lab course at the Institute for Communications Technology (IfN), TU~Braunschweig. The goal of the experiment is to provide students with an intuitive and reproducible introduction to MC concepts using a low-cost and accessible fluidic setup. The system employs a background water flow into which three dye colors are injected and symbols are detected by a multi-wavelength photosensor. A zero-forcing--based estimator is used to separate the spectral components and reliably identify the transmitted colors.
The experiment is designed to be completed independently by students within a single laboratory session and requires only basic prior knowledge from introductory communication engineering courses. A detailed script accompanies the experiment, guiding students through channel characterization, color detection, pseudoinverse computation, and simple data transmission using on-off keying. In pilot trials, students successfully reproduced the entire communication chain and achieved stable data rates of up to 0.5~bit/s over a 15~cm channel.
The proposed testbed demonstrates that fundamental principles of MC can be taught effectively using a compact and inexpensive experimental setup. The experiment will be integrated into an undergraduate lab course.
\end{abstract}
\section{Introduction}

\Ac{MC} has emerged as a promising paradigm for future communication systems, particularly in environments where conventional electromagnetic, acoustic, or optical signaling is ineffective or impractical. Such environments include biological tissues, confined industrial spaces, and nano- to microscale systems where fundamental physical constraints limit the applicability of traditional communication technologies. \ac{MC} leverages chemical messengers as carriers of information, enabling biocompatible and energy-efficient signaling that mirrors processes widely found in nature~\cite{akyildiz2019molecular, nakano2013molecular}.

Natural molecular processes, ranging from pheromone-based signaling in insects~\cite{unluturk2017plant} to bacterial quorum sensing and inter-species communication~\cite{boo2024microbe}, illustrate the versatility and functional richness of molecule-based information exchange. Beyond biological systems, \ac{MC} concepts are increasingly explored for technological applications. Proposed use cases include monitoring chemical reactions and nanoscale manufacturing processes~\cite{jamali2019tutorial}, automated quality control in the food and beverage sector via electronic olfactory sensing~\cite{vanarse2022malt}, immersive olfactory communication interfaces for consumer electronics~\cite{giannoukos2017mass}, and robust communication links in harsh or electromagnetically shielded environments such as mining tunnels~\cite{qiu2014tunnels}.

The potential impact of \ac{MC} is especially pronounced in biomedical engineering. Due to its inherent biocompatibility and event-driven nature, \ac{MC} is viewed as a key enabling technology for tasks such as biomarker detection, targeted drug delivery, immune-system monitoring, and future nanoelectronic interfaces for the Internet of Bio-Nano Things~\cite{kuscu2021ibnt, dissanayake2021performance}. Research in this area aims to bridge biological communication pathways with external networks through bio-cyber interfaces and next-generation 6G systems~\cite{6G-life, hofmann2022mcnfv}.

Despite the breadth of ongoing research, \ac{MC} remains largely absent from standard undergraduate curricula in electrical and information engineering. While \ac{MC} is covered in select graduate-level seminars or specialized programs, students rarely have the opportunity to gain hands-on experience with physical implementations. This lack of practical exposure makes it difficult for students to connect theoretical channel models, such as diffusion dynamics, Brownian motion, and chemical reactions, to real-world communication constraints.

To address this gap, we present a simplified, low-cost, and reproducible \ac{MC} testbed designed specifically for undergraduate laboratory education. The testbed is based on the platform presented by Wietfeld et al.~\cite{wietfeld2025evaluation} and uses dye-based signaling transported by laminar water flow and detected using a multi-wavelength photosensor. Despite its simplicity, the setup captures essential \ac{MC} phenomena, including flow-driven advection, pulse spreading, inter-symbol interference, and spectral separation through a \ac{ZF} \ac{MIMO}-inspired approach.

The experiment was developed for the fourth-semester \textit{Communication Engineering} lab course at the Institute for Communications Technology (IfN) of TU~Braunschweig in Germany. Particular emphasis was placed on enabling students to conduct the experiment autonomously after studying an accompanying script. The experiment requires only foundational knowledge from introductory courses yet it exposes students to the structure of a full communication chain: channel characterization, transmitter design, signal detection, equalization, and message decoding.

Pilot trials demonstrated that students were able to reproduce all major steps within a single laboratory session and obtained interpretable and meaningful results. Therefore, the experiment provides an accessible and engaging introduction to \ac{MC}, complementing traditional electromagnetic-focused coursework.

The contribution of this work is threefold:
\begin{itemize}
    \item We adapt a compact, robust, and low-cost \ac{MC} testbed~\cite{wietfeld2025evaluation} for undergraduate education.
    \item We develop a detailed, self-contained instructional script that guides students from theoretical foundations to experimental execution and data analysis.
    \item We validate the testbed in a pilot run, demonstrating reproducibility, student comprehension, and realistic transmission performance.
\end{itemize}

Beginning in the winter term 2025/2026, the proposed experiment \emph{``A Testbed for Molecular Communication''} will be used in the \textit{Communication Engineering} lab course, providing students with practical insight into a rapidly growing research field with significant implications for future nano-, biomedical-, and 6G-enabled communication technologies.

\section{Fundamentals}
\label{sec:grundlagen}

\subsection{Types of Molecular Communication}

To implement \ac{MC}, several approaches exist. They differ primarily in whether the information-carrying molecules are released directly into the medium or transported from the transmitter to the receiver by some form of ``motor''.

\begin{enumerate}
    \item Diffusion: Diffusion-based communication is a principle that occurs naturally in the human body. In this case, molecules or biochemical messengers are released by a transmitter into a medium such as blood, where they propagate by passive means such as Brownian motion or pre-existing flow systems until they are absorbed by a receiver~\cite{diffusion}.
    \item Gap Junction Channels: Another mechanism found in nature that can be used for \ac{MC} is the gap junction channel. In this form of communication, molecules are exchanged between adjacent cells through natural channels that directly connect their cytosol~\cite{jamali2019tutorial}.
    \item Molecular Motors: The molecular motor approach is based on the idea that one or more molecules are carried inside a motor protein that moves along a filament-like track. This movement is achieved by repeatedly binding to and detaching from the track, enabling stepwise transport toward the receiver~\cite{mallik2004motors}.
    \item Bacterial Transport: In nature, bacteria move using flagella, i.e. protein filaments that rotate like propellers. This type of locomotion can be exploited for bacterial-transport-based \ac{MC}. Here, the information molecules are encapsulated within an engineered bacterium, which is guided toward the receiver using a nutrient source~\cite{bakterien_transport}.
\end{enumerate}

Each of these \ac{MC} strategies has its own advantages and limitations. A central question is: Which mechanism is suitable for nanorobots on the scale of blood platelets, especially for potential applications inside the human body? Key factors include the extremely limited space available for implementation as well as the minimal energy resources at the nanoscale~\cite{gohari2016information}.

Molecular motors, gap junction channels, and artificial bacteria have the disadvantage that they require additional supporting infrastructure, which is often impractical. Moreover, molecular motors consume too much energy for use in nanoscale devices.

Diffusion-based communication is the simplest to implement, since molecules only need to be released into the surrounding medium and no additional structure is required. However, diffusion is entirely uncontrolled: molecules experience random propagation delays, resulting in highly variable arrival times at the receiver. This can cause channel distortion and lead to information loss~\cite{nakano2013molecular}.

\subsection{Diffusion-Based Communication Channel}
\label{subsec:diffusion_channel}

By choosing diffusion-based \ac{MC}, several characteristics of the communication channel must be taken into account. One key aspect is the stochastic movement of molecules, which is governed by Brownian motion. As a result, the spatial distribution of molecules can be modeled as statistically normally distributed~\cite{gohari2016information}.

\subsubsection{Statistical Distribution in the Channel}

In a medium, molecules propagate according to concentration gradients, which can be described by Fick's laws of diffusion on a macroscopic scale. The first Fick's law in one spatial dimension relates the particle flux density \( J \), the diffusion coefficient \( D \), and the concentration gradient \( \frac{\partial \rho}{\partial x} \) as follows:
\begin{equation} \label{eq:2.1}
    J(x,t) = -D \frac{\partial \rho(x,t)}{\partial x}.
\end{equation}
Combining this with the conservation of mass and incorporating the molecular production density \( c(x,t) \) at position \( x \) yields
\begin{equation} \label{eq:2.2}
    \frac{\partial \rho(x,t)}{\partial t} 
    = - \frac{\partial J(x,t)}{\partial x} + c(x,t).
\end{equation}
Substituting \eqref{eq:2.2} into \eqref{eq:2.1} results in the second Fick's law, extended by the molecular production term:
\begin{equation} \label{eq:2.3}
    \frac{\partial \rho(x,t)}{\partial t}
    = D \frac{\partial^2 \rho(x,t)}{\partial x^2} + c(x,t).
\end{equation}
To solve this differential equation, we assume that the domain \( I = (-\infty,\infty) \) has no reflective or absorptive boundaries. A transmitter located at \mbox{$x = 0$} releases a molecule at time \( t = 0 \), i.e., the source term is \( c(x,t) = \delta(x=0)\delta(t=0) \). Under these assumptions, the solution becomes
\begin{equation} \label{eq:2.4}
    \rho(x,t) 
    = \frac{\mathbf{1}[t>0]}{(4\pi Dt)^{0.5}}
      e^{-\frac{x^2}{4Dt}}.
\end{equation}
For any fixed time \( t \), this solution is a Gaussian distribution in \( x \). It should be noted that, in practice, external influences such as chemical reactions or absorption may prevent a closed-form analytical solution.

On a microscopic level, each molecule undergoes Brownian motion, which can be modeled as a random walk process. In the one-dimensional case, each molecular displacement is independent of previous steps and can be treated as a normally distributed random variable. Since the sum of normally distributed variables is also normal, it follows that for \( 0 \le t_1 < t_2 \), a molecule released at time \( t=0 \) at the origin is normally distributed as  
\[
\mathcal{N}(0, 2D(t_2 - t_1)).
\]
Hence, a molecule located at position \( x \) at time \( t \) exhibits a variance of \( 2Dt \), and the resulting distribution matches the form in \eqref{eq:2.4}. This agrees with Einstein’s classical 1905 result linking Brownian motion and Fick’s laws~\cite{gohari2016information,ficksches_Gesetz}.

\subsection{Receiver Models}

Two principal receiver types are considered in diffusion-based \ac{MC}: those that do not alter the molecular concentration in the medium, and those that absorb or chemically react with the molecules.

\subsubsection{Non-Reactive Receivers}

Non-reactive receivers include the \emph{sampling receiver} and the \emph{transparent receiver}.  
The sampling receiver measures the molecule concentration at a specific point in space, while the transparent receiver measures the concentration within a spherical volume \( V_R \).

Both are assumed to perfectly detect the number of molecules. The main difference lies in their output:  
the sampling receiver measures the mean concentration \( \rho(x,t) \), whereas the transparent receiver measures both the mean \( \rho(x,t) \) and the variance \( \rho(x,t)/V_R \).

Since non-reactive receivers do not remove molecules from the environment, the concentration remains unchanged and the Fick diffusion model without boundary conditions can be applied~\cite{gohari2016information}.

\subsubsection{Reactive Receivers}

Reactive receivers include the \emph{absorbing receiver}, which removes molecules from the environment upon arrival, thereby altering the local concentration. Another important model is the \emph{ligand-based (reactive) receiver}, analogous to cellular receptors. Molecules bind to receptor sites on the surface, which may be in either a ``bound'' or ``unbound'' state. Molecules cannot bind to receptors that are already occupied, and receptor unbinding occurs according to a stochastic temporal process.

The number of bound receptors can be modeled by a memoryless binomial distribution \(\mathrm{Binomial}(k,p)\), where \( k \) is the number of receptors and \( p \) is the binding probability.

As a result, the received signal can be modeled as
\[
Z = \alpha X + N,
\]
where \( Z \) is the receiver output, \( \alpha \) a gain factor, \( X \) the molecule concentration, and \( N \) Gaussian noise.

To apply Fick's laws to both reactive and non-reactive receivers, suitable boundary conditions must be imposed: absorbing receivers correspond to zero-boundary surfaces, whereas ligand-based receivers correspond to partially absorbing boundaries with active molecular interaction~\cite{gohari2016information}.

\subsection{Molecules}
\label{subsec:molecules}

The design of an \ac{MC} system also requires specifying the types of molecules used for signal transmission and measurement. Several experimental approaches have already been investigated in the literature. A key distinction is made between \ac{SIMO} communication, where communication relies on one molecular species, and \ac{MUMO} communication, where several molecular types are used simultaneously for data transmission~\cite{wietfeld2025evaluation}.

\subsubsection{Single-Molecule Communication}

One approach to \ac{SIMO} uses acids and bases as information carriers. In these experiments, acids or bases are released and transported by several peristaltic pumps toward the receiver, typically a pH probe. Using recurrent neural network detectors, data rates of up to 4~\(\frac{\mathrm{bit}}{\mathrm{s}}\) have been achieved~\cite{MC_Base_Acid}.  

Another \ac{SIMO} method employs switchable fluorescent proteins, whose key advantage lies in their ability to be reused for multiple transmission cycles. This setup requires two LED arrays at the transmitter for activating and deactivating the proteins. At the receiver, a spectrometer is used to measure fluorescence intensity, achieving data rates of approximately 0.1~\(\frac{\mathrm{bit}}{\mathrm{s}}\)~\cite{floureszierendeProteine}.  

These two approaches have so far been demonstrated only over communication distances of a few centimeters. For longer-range communication, magnetic nanoparticles have been released by peristaltic pumps and detected using a non-invasive susceptometer. Over a short distance of 5~cm, these magnetic nanoparticles enable data rates of 6~\(\frac{\mathrm{bit}}{\mathrm{s}}\), and messages remain decodable even over distances of up to 40~cm~\cite{MC_magneticParticels}.

\subsubsection{Multi-Molecule Communication}

In \ac{MUMO}, several molecular species are used for data transmission. One straightforward choice is the use of different colors as distinct molecular types~\cite{wietfeld2025evaluation}.

In the experimental part of this work, this principle is applied using multiple dyes. For decoding, a linear estimator is used that determines the time-dependent color intensities from the raw spectral measurements of the photosensor for each color employed. Using this method, data rates of approximately 3~\(\frac{\mathrm{bit}}{\mathrm{s}}\) have been demonstrated over distances greater than 20~cm~\cite{wietfeld2025evaluation}.

\section{Fundamentals of the Laboratory Experiment}
\label{sec:praktikumgrundlagen}
\subsection{Communication Channel}
\label{subsec:kanal}

For the proper execution and interpretation of the laboratory experiment, it is essential to understand the characteristics of the communication channel—here, a water-filled tube—and in particular the behavior of the dye molecules within it. The transport of the dye molecules through the channel is dominated by \emph{advection}, meaning that the molecules are carried from the transmitter to the receiver by the background flow of the water~\cite{wietfeld2025evaluation}.  

In fluid-flow–based \ac{MC} systems, both the nature of the flow and the residual diffusion of the molecules must be considered in order to predict and analyze their behavior within the medium.

\subsubsection{Flow Regimes}

Fluid motion is typically classified into \emph{laminar} and \emph{turbulent} flow. Laminar flow is characterized by smooth, orderly movement without vortices, whereas turbulent flow involves chaotic, swirling motion. The classification is determined by the Reynolds number \([Re]\), given by
\begin{equation} \label{eq:2.5}
    Re = \frac{d_c \, v_{\text{avg}}}{\nu},
\end{equation}
where \( d_c \) is the channel diameter, \( \nu \) is the kinematic viscosity of the fluid (for water \( \nu \approx 1.01 \times 10^{-6} \, \mathrm{m^2/s} \)), and  
\[
v_{\text{avg}} = \frac{Q_0}{\pi r_c^2}
\]
is the average flow velocity, with \(Q_0\) denoting the background flow rate.

The transition from laminar to turbulent flow occurs around \( Re = 2100 \).  
Thus, flow conditions with \( Re < 2100 \) can be treated as laminar.  
In the laboratory experiment, the flow inside the tube is well within the laminar regime, as we will calculate in Section~\ref{subsec:flow_calc} below.

For laminar flow in circular tubes, the velocity profile as a function of radial position \( \rho \) is given by the Poiseuille flow model:
\begin{equation} \label{eq:2.7a}
    v(\rho) = v_{\text{max}} \left(1 - \frac{\rho^2}{r_c^2}\right), \quad \rho \in [0, r_c],
\end{equation}
where the maximum flow velocity satisfies~\cite{wietfeld2025evaluation,white2016fluid}
\[
v_{\text{max}} = 2 \, v_{\text{avg}}.
\]

\subsubsection{Diffusion}

The influence of molecular diffusion relative to advection can be quantified using the P\'eclet number \([Pe]\):
\begin{equation} \label{eq:2.7b}
    Pe = \frac{r_c \, v_{\text{avg}}}{D},
\end{equation}
where \(D\) is the molecular diffusion coefficient.  
For water-based ink, the diffusion coefficient satisfies~\cite{wietfeld2025evaluation}
\[
D \lesssim 2.299 \times 10^{-9} \, \mathrm{m^2/s}.
\]

For the parameters used in the experiment, the resulting P\'eclet number is significantly greater than one, see Section~\ref{subsec:flow_calc}. This indicates that advection dominates over diffusion, meaning that the background flow determines the molecule transport. Consequently, diffusion can safely be neglected in the practical execution of the experiment~\cite{wietfeld2025evaluation,white2016fluid}.
\subsection{Determination of Colors in the Channel}
\label{sec:color_detection}

In the laboratory experiment, three different dye colors are injected in a controlled manner at the transmitter. These colors must be individually detectable at the receiver. To achieve this, a \ac{ZF} approach is applied.

In classical wireless communications, \ac{ZF} assumes a system with \( M_t \) transmit antennas and \( M_r \) receive antennas. For the laboratory experiment, this concept can be applied analogously:  
the different injected dye colors correspond to the transmit dimensions \( M_t \), and the measurable wavelength intensities of the photosensor at the receiver correspond to the receive dimensions \( M_r \).

Thus, a channel matrix \( H \in \mathbb{C}^{M_r \times M_t} \) can be defined for the transmission.  
To determine the contributing color components using \ac{ZF}, a MIMO equalization matrix  
\( A \in \mathbb{C}^{M_t \times M_r} \) is required.  
The received signal can then be written as
\[
y = H x + N,
\]
where \( N \) represents noise.  
Applying the \ac{ZF} equalizer yields
\[
\tilde{x} = A y, \qquad \tilde{x} \in \mathbb{C}^{M_t}.
\]

Each component of the estimated vector \( \tilde{x} \) is then mapped to the closest valid input symbol through
\begin{equation} \label{eq:2.8}
    \hat{x}_i = 
    \arg\min_{s \in X} \left| \tilde{x}_i - s \right|
    \quad \text{for all } i,
\end{equation}
resulting in the reconstructed input vector \( \hat{x} \).

To translate this theoretical method to the experiment, the channel matrix must first be determined for all dye colors. The simplest approach is to measure the impulse response for each color at the receiver. This yields  
\( H \in \mathbb{C}^{8 \times 3} \), for our specific setup, where the equalizer matrix \( A \) must be computed.  
If \(H\) is square, \(A\) can be obtained directly via matrix inversion.  
For non-square matrices, however, the Moore–Penrose pseudoinverse must be used:
\begin{equation} \label{eq:2.9}
    A = H^\dagger = (H^H H)^{-1} H^H,
\end{equation}
where \(H^H\) denotes the Hermitian transpose of \(H\).

Applying the equalizer then yields
\begin{equation} \label{eq:2.10}
    \tilde{x} = A y.
\end{equation}

Because the received vector \(y\) contains noise, \ac{ZF} generally amplifies this noise and \ac{ZF} is in general sub-optimal. 
However, in this experiment, the noise level is negligible because the signal strength—proportional to the color intensities—is significantly larger than the background noise.  
Finally, thresholding is applied to the components of \( \tilde{x} \) to detect the individual dye colors~\cite{goldsmith2005wireless}.
\section{Laboratory Experiment}
\label{sec:praktikum}

\subsection{Requirements}
\label{subsec:anforderungen}

The experiment developed in this work, i.e. an educational testbed for \ac{MC}, is intended for use in the laboratory course \emph{Communication Engineering}.  
The aim of this course is to deepen the student's knowledge in the fields of electrical engineering and information technology through a series of practical experiments. To achieve this, students are expected to familiarize themselves with the respective topics using an accompanying script before carrying out the experiments independently.

The course is designed for students in their fourth semester.  
Accordingly, the experiment and the corresponding script were developed such that the fundamental concepts of \ac{MC} can be understood even without advanced prior knowledge in communication theory. Nevertheless, basic understanding of data transmission, such as taught in the module \emph{Digital Communications}, is beneficial for fully grasping the underlying principles.

Since the laboratory course emphasizes hands-on learning, the experiment is designed to allow students to perform as many steps as possible independently. The concepts relevant to the experiment should be conveyed in a clear and intuitive manner, enabling students to directly observe and analyze the physical effects involved in \ac{MC}.

As research in general, particularly in biomedical and nanoelectronics contexts, continues to advance rapidly, future applications inside the human body are expected to rely on molecule-based communication. Therefore, this laboratory experiment  provides students with an initial introduction to \ac{MC}, offering insight into a communication paradigm that is likely to play a central role in emerging nanoscale systems.

\subsection{Objectives of the Experiment}
\label{subsec:ziele}

At present, the Institute of Communications Technology offers no dedicated lecture focusing specifically on \ac{MC}. Therefore, when designing an appropriate laboratory experiment, it must be ensured that only general prior knowledge, such as channel impulse responses and digital communication principles taught in the third-semester course \emph{Signals and Systems}, is required.

For this reason, the experiment must be structured such that message transmission remains easily understandable. This includes using molecules that are visible to the naked eye and choosing modulation or coding schemes that are not overly complex. The overarching goal is to demonstrate to students that, in addition to conventional antenna-based communication methods, alternative techniques exist—such as transmitting short messages using \ac{MC} with colored molecules.  

The experiment is designed to guide students, step by step, toward assembling the communication channel themselves. To support this, a dedicated experiment script was developed. 

The first chapter introduces \ac{MC}, explains its relevance, and highlights potential application areas.  
The second chapter familiarizes students with the different types of \ac{MC}, with a particular focus on diffusion-based systems. It also introduces several mathematical expressions necessary for describing the system and concludes with an overview of receiver types and suitable molecular candidates.

Chapter three covers the fundamental concepts required for the experiment, including fluid-dynamic principles such as flow regimes, as well as the \ac{ZF} approach from wireless communications.

Chapter four provides a detailed description of the experimental setup and components. It also includes important notes regarding programming, safety, and careful handling of laboratory equipment.

Chapter five contains the instructions for the experiments to be carried out during the laboratory session, along with protocol templates that students must complete during the experiment.

The final chapter includes preparatory homework assignments that students are required to work through before attending the laboratory session. These tasks ensure that participants arrive with a sufficient foundational understanding.

\subsection{Challenges and Solutions in Developing the Experiment}
\label{subsec:probleme}

During the development of the \ac{MC} experiment, several technical and practical challenges emerged. This section discusses the issues encountered and the approaches used to resolve them.

\subsubsection{Pumps}

At the start of the experiment design, the main pump was unable to draw water upward from the reservoir against gravity. This issue was resolved by constructing a 3D-printed water container that allowed the pump to utilize gravity when generating the background flow.

When the setup was later adapted to support dye injection via Y-connectors, another issue arose: the dye pumps remained permeable even when switched off. Shortly after assembly, dye was observed flowing backward into the background-flow tubing. This revealed that the pumps were not sealing the lines, allowing liquid to flow according to height differences. The problem was solved by repositioning the containers, i.e. placing the dye reservoirs at the lowest point and the waste container at the highest point—thus preventing unintended flow of the background liquid.

\subsubsection{Y-Connectors}

Initially, the use of Y-connectors to merge dye tubing with the background flow yielded promising measurement results. However, a major issue soon became apparent: vortices formed inside the connectors, causing background water to push dye upward into the dye tubing. As a result, during subsequent injections, the pumps first expelled this back-flowed water instead of dye, delaying the intended injection.

The first attempted solution was to eliminate Y-connectors entirely and insert small injection needles directly into the background-flow tubing, one for each dye. While this produced very clean dye pulses for a few milliseconds, it was not a practical long-term solution. Correct insertion of the needles was difficult, the tubing could easily be punctured, and the needles could slip out during operation. Once a needle slipped or pierced the tubing incorrectly, the segment became unusable due to uncontrolled leakage, even when the pumps were off.

Because Y-connectors provided a reliable physical interface, the next idea was to glue the injection needles into the Y-connectors using epoxy resin. While this reduced back-flow slightly, it did not eliminate the problem.

A subsequent modification introduced a small baseline voltage applied to the pumps to keep them slightly active. However, because each pump had a different height relative to the main flow pump, they required different baseline voltages. Early experiments with this approach still showed mixing of dye and water near the Y-connectors.

Ultimately, the issue was fully resolved by installing check valves between the dye pumps and the injection needles fixed in the Y-connectors. After adding the check valves, no back-flow occurred and the injected dye volume became reproducible and stable.

\subsubsection{Photosensor}

The receiver’s photosensor was initially illuminated by a white LED with a 100~\(\Omega\) series resistor, producing a very high light intensity. This led to saturation in the measured wavelength intensities. The resistor was therefore increased to 9.5~k\(\Omega\), reducing the LED brightness to a suitable level.

Moreover, the photosensor was originally exposed to ambient light, causing significant fluctuations in the measurements over time. This problem was addressed in two ways:  
(1) the sensor and LED were enclosed in a 3D-printed housing with openings only for the tubing and cables;  
(2) the software initially averaged intensity readings over a 10-second window to detect deviations.

However, this averaging approach was not sufficiently robust for magenta and yellow dyes, whose absorption spectra are very similar. Consequently, the system was resigned using the \ac{ZF} method from wireless communications: by applying the pseudoinverse of the channel matrix to the output vector, an estimated input vector is obtained. Thresholding this vector yields the detected color. This method is robust enough that averaging is no longer needed and avoids unnecessary computational overhead.

\subsubsection{Message Transmission}

\Ac{OOK} was used for bit transmission. In each 3-bit symbol, cyan represents the most significant bit and yellow the least significant bit. At the receiver, each predetermined time slot is inspected, and a detected color sets the corresponding bit to 1 in the sequence \((\text{cyan}, \text{magenta}, \text{yellow})\).

A major challenge in reliable message transmission was synchronization between the transmitter and receiver. Initially, a preamble consisting of eight consecutive ones was sent, but this approach was too slow and highly susceptible to bit errors. Instead, an invalid bit using the yellow dye located closest to the transmitter and, thus, having the shortest channel delay, was adopted as a start and stop flag. The yellow dye is injected in a larger volume to produce a clearly detectable peak.

Early transmissions revealed frequent misalignment between color arrivals and their assigned time slots. To resolve this, the receiver was set to wait slightly less time than the transmitter after the start flag. Specifically, for transmitter wait time \(t_{\text{start}}\) and slot duration \(t_{\text{bit}}\), the receiver waits  
\[
t_{\text{Receiver}} = t_{\text{start}} - \frac{t_{\text{bit}}}{2},
\]
while the transmitter uses  
\[
t_{\text{Sender}} = t_{\text{start}}.
\]
This ensures that colors arrive approximately in the center of each time slot.

Another problem occurred during longer transmissions: the processing time of received signals varied, causing the internal system clock to drift by up to 300~ms. This eventually desynchronized the bit sequences. The solution was to compute all future time slot boundaries relative to the initial start-time reference. After transmitting each symbol, its duration was added to the original start timestamp, preventing cumulative timing drift.

\subsection{Instructions for Conducting the Experiment}
\label{sec:durchfuehrung}

\noindent\textbf{Homework:} The laboratory session should begin with a short colloquium. Students are required to complete the assigned homework in advance at home. The preparatory tasks are as follows:

\begin{enumerate}
    \item Determine the flow regime and the Péclet number for a background flow of \(Q_0 = 10~\mathrm{ml/min}\) and a channel diameter of \(d_c = 2~\mathrm{mm}\).
    \item Provide a fundamental explanation of the \ac{ZF} method.
\end{enumerate}

These tasks, along with several additional conceptual questions, should be discussed during the colloquium.  

\noindent\textbf{Colloquium:} A recommended structure for the colloquium begins with a set of introductory questions to ease students into the topic. Possible questions include:

\begin{enumerate}
    \item \textbf{Why is \ac{MC} needed?}  
    \Ac{MC} is essential for applications at nanoscales, such as within the human body, where energy resources are extremely limited and electromagnetic waves experience high propagation losses, making traditional communication infeasible~\cite{gohari2016information}.

    \item \textbf{Explain the different types of \ac{MC}.}  
    Students may be given the blank schematic from the appendix and asked to label and explain the four main types, following the discussion from Chapter~2.

    \item \textbf{Which flow regimes exist, and how are they determined?}  
    This question connects directly to the homework. The expected solution is:  
    \[
    v_{\text{avg}} \approx 0.053 \,\frac{\mathrm{m}}{\mathrm{s}}, \qquad 
    Re \approx 104.95,\qquad 
    Pe \approx 23053.5.
    \]
    The flow is laminar because \(104.95 < 2100\), and diffusion can be neglected since \(23053.5 \gg 1\).

    \item \textbf{What influence does the flow regime have on molecular movement?}  
    In turbulent flow, particles experience movement not only in the flow direction but also orthogonally, causing interruptions in the continuous dye front. The measured intensity would show recurring peaks rather than the smooth increase and decrease characteristic of laminar flow.

    \item \textbf{Explain the \ac{ZF} method used in the experiment.}  
    Because the topic is complex, students should prepare this beforehand. In essence, \ac{ZF} uses the known channel matrix to infer how a transmitted signal appears at the receiver. By applying the inverse or pseudoinverse of the channel matrix, the receiver reconstructs the transmitted symbol and can thereby distinguish the dye colors.

    \item \textbf{How are the pumps and the photosensor controlled during the experiment?}  
    Students must understand how to control the pumps using the interface \texttt{(Pumpe1, Pumpe2 Dauer1, Dauer2)}.  
    The background flow is always started with a \texttt{1} and stopped with a \texttt{0}.  
    The photosensor must always be calibrated using a clean water-filled tube; calibration is initiated by sending a \texttt{1} via the serial monitor. A new calibration is required after each firmware upload.
\end{enumerate}

Only after completing the colloquium and demonstrating sufficient understanding of the theoretical and practical foundations may students proceed to the experimental setup.

\subsection{Setup}
\label{subsec:aufbau}

During the setup phase, all participants must be instructed to wear gloves and laboratory coats to prevent accidental contact with dyes on skin or clothing. While protective clothing is not strictly necessary for students operating the computer during the experiment, it is mandatory during the physical setup, where the risk of spilled dye is higher.

A critical point during setup is that the tubing must \emph{not} be filled while the pumps are connected to the system, as this may cause damage to the pump mechanisms. Because only small quantities of dye are required for the experiment, the dye containers should be filled with no more than 1~ml per color to avoid unnecessary material consumption.

\subsection{Experiment Execution}
\label{subsec:durchfuehrung}

Throughout the experiment, the water reservoir supplying the background flow must contain a sufficient amount of water to ensure that the pump does not run dry. When manually activating the dye pumps, the activation duration should not exceed 100~ms, as longer pulses lead to unreliable measurements and unnecessary dye usage.

Participants must adhere to the correct input format  
\[
\texttt{(Pumpe1, Pumpe2\ Dauer1, Dauer2)},
\]  
and should be reminded of the proper control procedure for all pumps.  
If an error occurs while uploading the program code to the controller, the reset button on the ESP must be pressed. When viewing the device directly, this is the right-hand button next to the USB-C connector.

During the section of the experiment involving the computation of the pseudoinverse of the channel matrix, it is essential that students construct the matrix correctly: columns must be ordered yellow–magenta–cyan, and rows must be arranged in ascending order of wavelength.

In the final part of the experiment, when entering messages for transmission, students should choose short messages to avoid unnecessarily long waiting times caused by extended bit sequences.

\subsection{Pilot Experiment with Students}
\label{subsec:probeversuch}

Before the experiment was officially integrated into the student laboratory course, the complete procedure was carried out in a pilot trial with several students. The purpose of this preliminary run was to evaluate the clarity of the script, verify the comprehensibility of the underlying concepts, and assess whether the entire experiment could be completed within the intended time frame.

During the pilot experiment, it became evident that students must thoroughly study the laboratory script as well as the theoretical background in advance in order to fully grasp the complexity of the topic and successfully complete the experiment within the allotted time.  
Furthermore, it was observed that the required program commands had to be explained precisely at the beginning of the session. For this reason, the script was expanded to include an additional table containing example program commands to guide the students more effectively.

The total duration of the pilot experiment suggests that the finalized laboratory session will require approximately two and a half hours. The final part of the experiment, which focuses on maximizing the achievable data rate, offers a natural point of flexibility in the schedule: faster-working students can attempt to optimize the data rate, while slower-working students can complete the previous tasks at a comfortable pace and still obtain meaningful results that help them understand the complexity of \ac{MC}.

In summary, the pilot experiments demonstrated that the laboratory script in its current form is clear and well-structured. The students were largely able to carry out the tasks independently, indicating that the experiment is suitable for integration into the communication engineering laboratory course.
\section{Experiment}
\label{sec:experiment}

\subsection{Experimental Setup}
\label{subsec:versuchsaufbau}

\begin{figure}[h]
    \centering
    \includegraphics[width=0.4\textwidth]{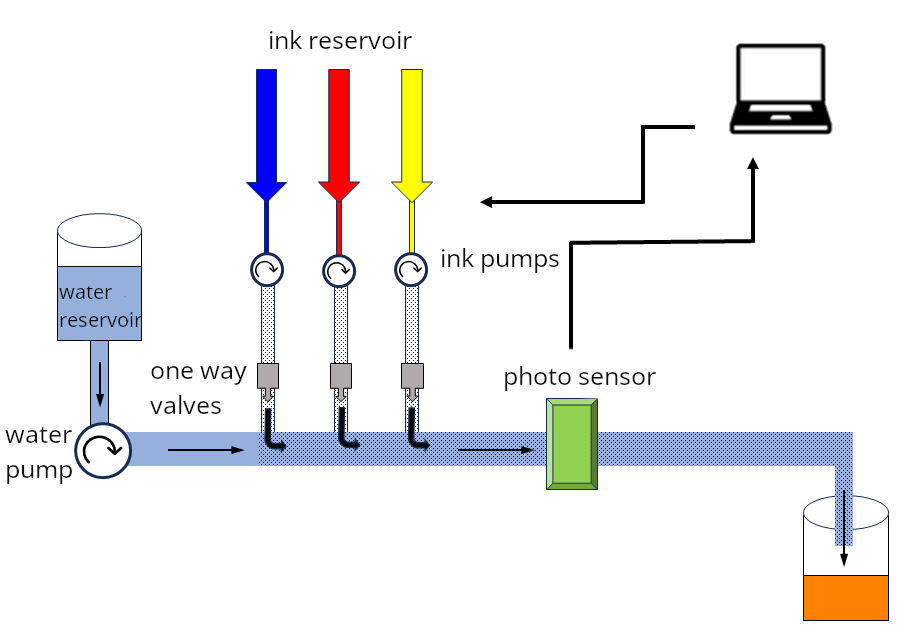}
    \caption{Schematic overview of the experimental setup}
    \label{fig:versuchsaufbau}
\end{figure}

In this experiment, water is pumped from a reservoir through a tube using a membrane pump, thereby generating a background flow. Into this flow, dye in the colors cyan, magenta, and yellow can be injected using three additional pumps. The injection is realized via Y-connectors into which hypodermic needles have been glued using epoxy resin. Integrated check valves prevent water–dye mixtures from flowing back into the dye supply lines. The resulting water–dye mixture then passes a photosensor capable of detecting individual wavelengths and assigning them to the corresponding dye colors using a dedicated software routine. Finally, the mixture flows into a waste container for later disposal. See Figure~\ref{fig:versuchsaufbau} for a schematic drawing of the entire setup.

\subsection{Pumps}

\begin{figure}[htbp]
    \centering
    \includegraphics[width=0.4\textwidth]{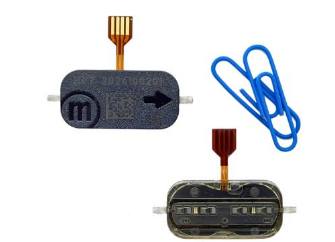}
    \caption{\textit{Bartels BP7} (Tubing) Pump}
    \label{fig:pumpe}
\end{figure}

Piezoelectric membrane pumps of the type \textit{Bartels BP7}, as shown in Figure~\ref{fig:pumpe}, are used for fluid transport. Piezoelectric materials deform when an electrical voltage is applied; this deformation is exploited in such pumps to transport liquids or gases with pressures of up to 500\,mbar. For safe operation, voltages outside the range of 0–250\,V must not be applied. For liquid transport, a modulation frequency of 100\,Hz is sufficient, while higher frequencies are required only for gas transport.

It is crucial to emphasize that the pumps are very sensitive to pressure. Their burst pressure is 1.5\,bar, and exceeding this limit will irreversibly damage them. Since syringes are used to store dye and to fill tubing prior to operation, students must not fill the tubing with fluid while the pumps are connected. Doing so may subject the pumps to excessive pressure and cause permanent damage~\cite{bartels2024}. The pumps are controlled via an ESP32 microcontroller mounted on the \textit{Bartels mp-Multiboard2}.

\subsection{Photosensor}

\begin{figure}[h]
    \centering
    \includegraphics[width=0.7\textwidth]{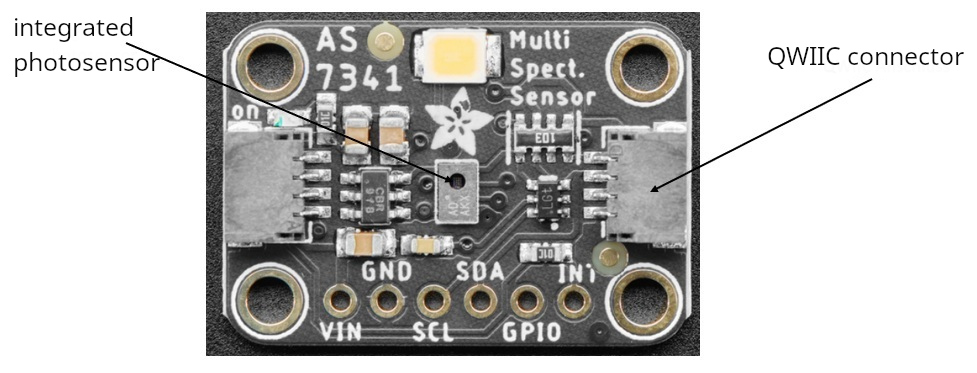}
    \caption{Adafruit AS7341 10-Channel Light Sensor}
    \label{fig:photosensor}
\end{figure}

The photosensor used in the experiment is an Adafruit AS7341 10-Channel Light Sensor, as shown in Figure~\ref{fig:photosensor}, which is capable of measuring nine different wavelengths—eight in the visible spectrum (415\,nm, 445\,nm, 480\,nm, 515\,nm, 555\,nm, 590\,nm, 630\,nm, 680\,nm) and one in the near-infrared range. It also provides a channel for measuring the total incident light intensity. The sensor communicates with an Arduino Micro via an \(I^2C\) interface, which reads the measured values and forwards them to the connected computer~\cite{adafruit_as7341}.

To avoid ambient light interference, the photodiode is placed inside a custom 3D-printed housing. A white LED mounted above the photodiode provides constant illumination and is controlled by the Arduino Micro. The fluid tube is routed between the LED and the photodiode to ensure that the sensor captures the intensity changes caused by the injected dye. During the experiment, only the visible-wavelength channels are evaluated.

\subsection{Check Valves}

\begin{figure}[h]
    \centering
    \includegraphics[width=0.3\textwidth]{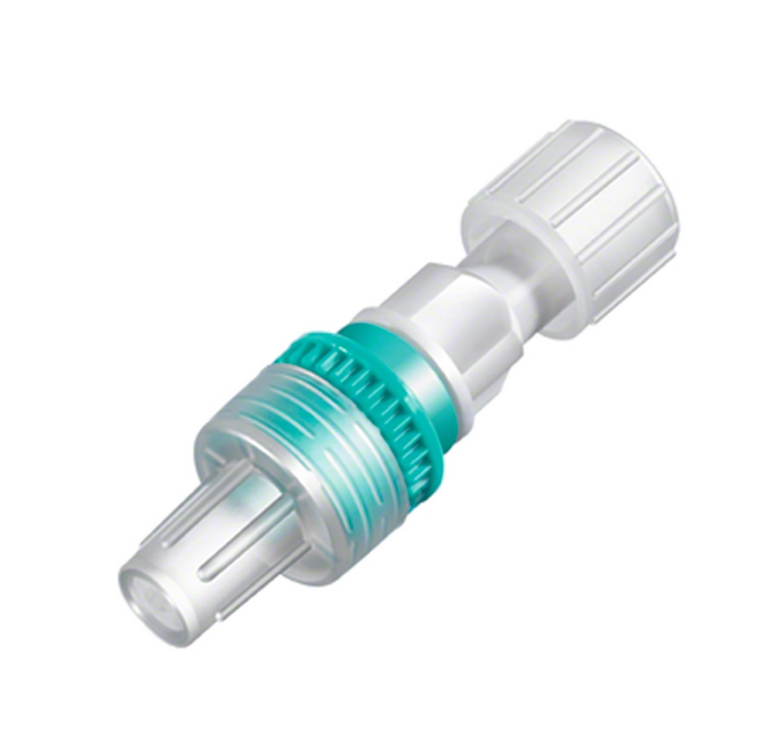}
    \caption{Infuvalve, B. Braun check valve~\cite{infuvalve_6bar}}
    \label{fig:checkvalve}
\end{figure}

Check valves are used to prevent water from flowing back into the dye supply lines and diluting the dyes. These valves, as shown in Figure~\ref{fig:checkvalve} allow flow only in one direction: from the white end to the green end. They open at a pressure of 20\,mbar. Since the pumps can generate pressures up to 500\,mbar, proper opening during dye injection is ensured. The valves seal reliably up to a back pressure of 2\,bar, which is never reached in the experiment. To avoid damage, pressures above 6\,bar must not be applied in the reverse direction. Consequently, the flow direction must be observed carefully during assembly~\cite{infuvalve_2bar,infuvalve_6bar}.

\subsection{Materials}

A complete list of components required for the experiment is provided in Table~\ref{tab:materials}.

\begin{table}[h]
\centering
\caption{Overview of materials used in the experiment}
\begin{tabular}{|l|l|l|}
\hline
\textbf{Category}       & \textbf{Components} & \textbf{Manufacturer/Part Number} \\ 
\hline
\multirow{9}{*}{Electronic Components} 
  & Arduino Micro & -- \\
  & Jumper cables & -- \\
  & Adafruit AS7341 Sensor & 4698 \\
  & mp-Multiboard2 & Bartels Mikrotechnik: BM-S-0008 \\
  & Micro-USB cable & Bartels Mikrotechnik: BM-S-0008 \\
  & Bartels BP7-Tubing Pump & Bartels Mikrotechnik: BM-S-0008 \\
  & mp-Highdriver4 & Bartels Mikrotechnik: BM-E-0003 \\
  & White LED & C513AWSNCX0Z0342 \\
  & 4.7\,k\(\Omega\) resistor & MFR100FTE734K7 \\
\hline
\multirow{7}{*}{Fluidic Components} 
  & Needles 30G x 1/2" & PT9969 \\
  & 3\,ml syringe & -- \\
  & TYGON LMT-55 tubing & Techlab: ISM SC0039T \\
  & Luer Lock adapter (male) & Techlab: UP P-850 \\
  & Luer Lock adapter (female) & Techlab: UP P-857 \\
  & Y-connectors 1/16" & -- \\
  & Infuvalve\texttrademark check valve & B. Braun PZN: 02232430 \\
\hline
\multirow{8}{*}{Miscellaneous} 
  & Cyan dye & Conrad: 2233365-VQ \\
  & Yellow dye & Conrad: 2233367-VQ \\
  & Magenta dye & Conrad: 2233366-VQ \\
  & Distilled water & -- \\
  & Waste container & -- \\
  & 3D-printed water reservoir & -- \\
  & Gloves & -- \\
  & Lab coat & -- \\
\hline
\end{tabular}
\label{tab:materials}
\end{table}

\subsection{Assembly Instructions}

To prevent damage to components, the assembly instructions must be followed precisely.  
The background flow section is assembled first. All tubing must be pre-filled with distilled water, as the pumps do not operate properly when air bubbles are present. Using a clean syringe equipped with a female Luer Lock adapter, water is pushed through each tube until a continuous water column is established. The filled tubes are then connected to the water reservoir, the Luer Lock adapter, and the background-flow pump (the flow direction is indicated by an arrow on the pump).

From this pump, one tube is connected to the cyan-dye Y-connector. Two additional Y-connectors are attached to include the magenta and yellow dye channels. The final Y-connector (yellow) is connected to a tube that runs through the photosensor housing and into the waste container.

\begin{figure}[h]
    \centering
    \includegraphics[width=0.7\textwidth]{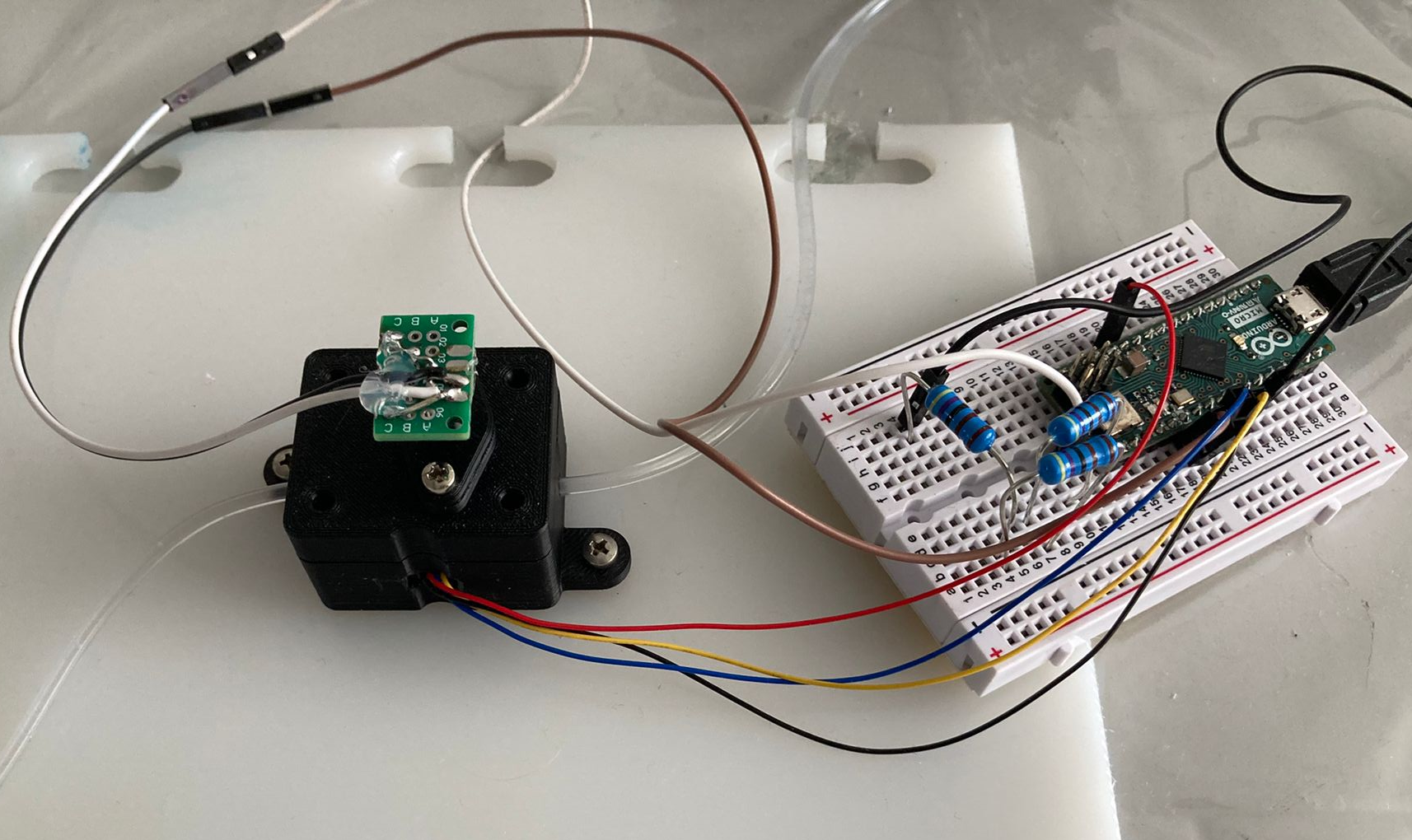}
    \caption{Experimental setup around the photosensor}
    \label{fig:photosensor_setup}
\end{figure}

Next, the dye-injection lines are assembled. Tubing from each dye pump is also pre-filled with water and connected to the pump outlet and the white end of a check valve. The epoxy-glued injection needle in each Y-connector is screwed onto the green end of the check valve. A male Luer Lock adapter is required to attach the valve. Pre-filling these tubes with water reduces the risk of spilling dye during assembly.

\begin{figure}[h]
    \centering
    \includegraphics[width=0.5\textwidth]{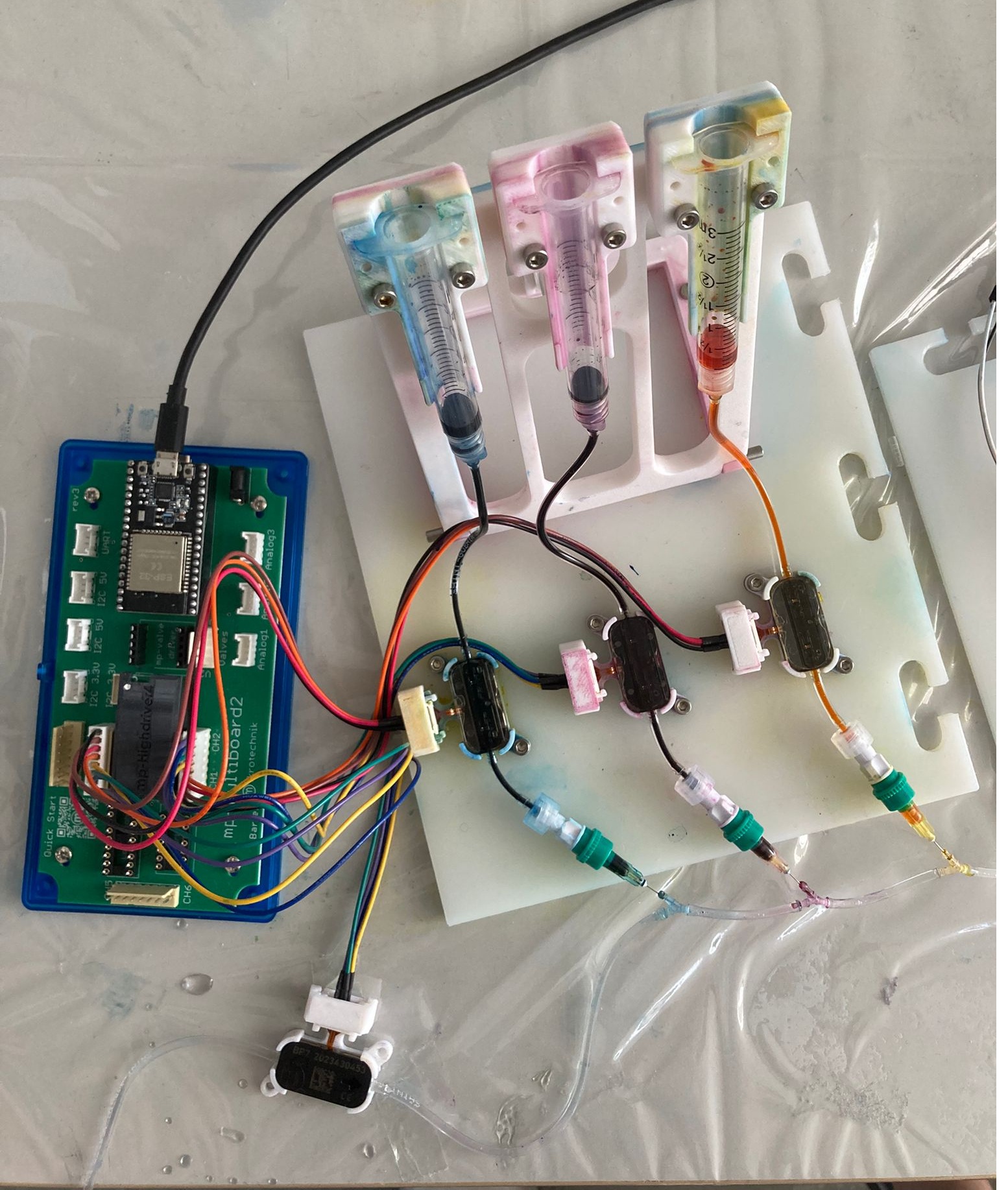}
    \caption{Assembly of the dye-injection system}
    \label{fig:pump_setup}
\end{figure}

Finally, the dye is added to the reservoirs, and both the ESP32 and the Arduino Micro are connected to the computer. Before beginning the experiment, any remaining water in the dye tubes must be pumped out until the dye reaches the Y-connectors. If dye accidentally enters the background-flow line, it can be flushed out using the background-flow pump.

\begin{figure}[h]
    \centering
    \includegraphics[width=0.5\textwidth]{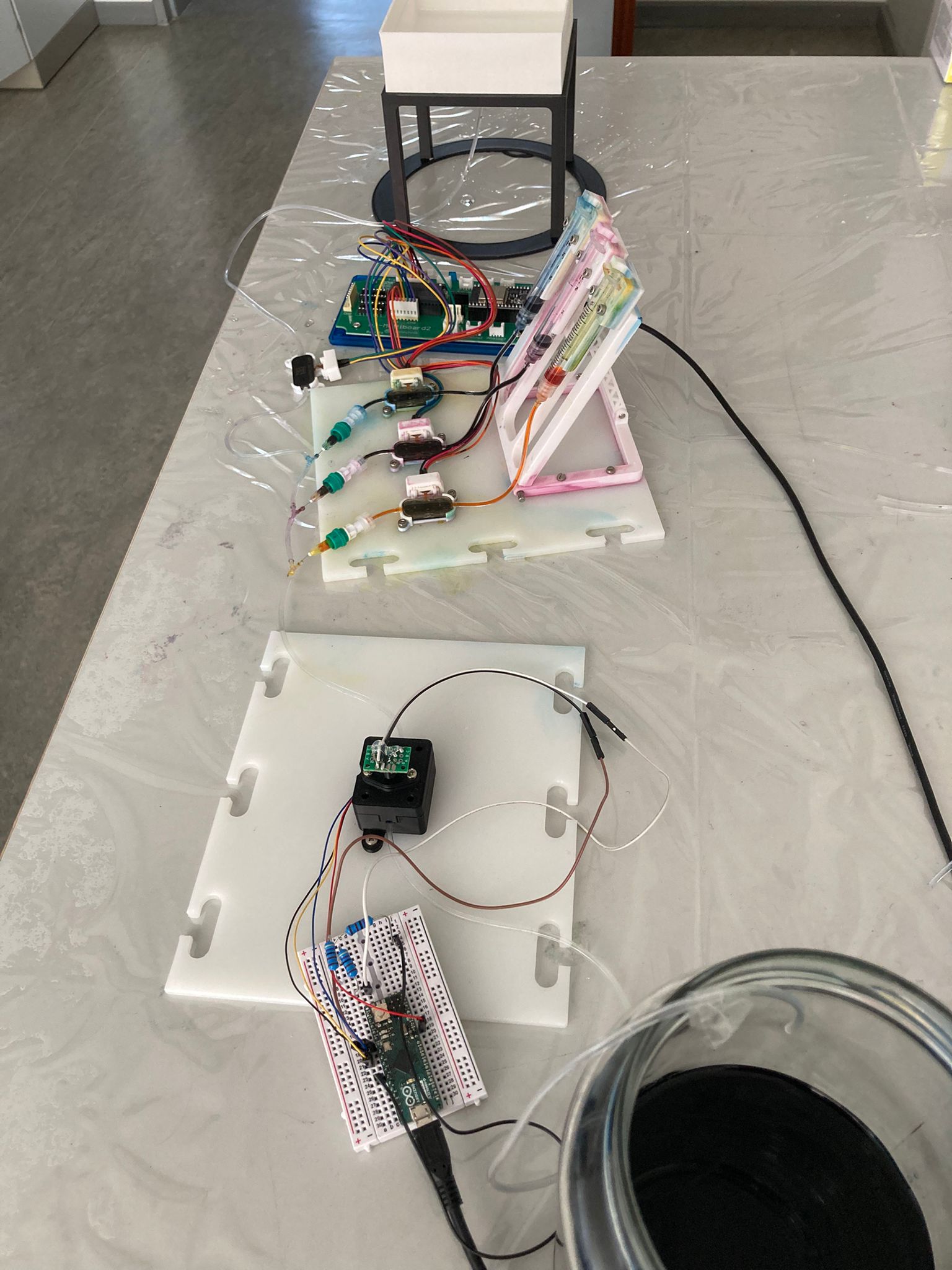}
    \caption{Complete experimental setup}
    \label{fig:complete_setup}
\end{figure}

Images of the finished assembly can be found in Figures~\ref{fig:photosensor_setup},  \ref{fig:pump_setup}, and \ref{fig:complete_setup} of the receiver side, transmitter side, and of the entire setup, respectively.

\subsection{Notes on Assembly}

It is strongly recommended that students wear gloves and lab coats during assembly to prevent dye from contacting skin or clothing. The positioning of containers is important because the pumps allow bidirectional flow when turned off. Although the check valves prevent water from entering the dye containers, they do not completely prevent dye from flowing into the background-flow tube under certain height differences. Therefore, the dye reservoirs must be placed lower than both the water inlet and the waste container. The waste container should be positioned highest, and the dye containers lowest, to avoid unintended dye flow that would affect sensor readings.

\subsection{Notes on Pump Control}

The dye pumps are numbered from right to left as Pump~1 to Pump~3 in the control software. Correct operation requires connecting each pump to the correspondingly labeled channel on the mp-Multiboard2, while the background-flow pump must be connected to Channel~4.

Control commands are entered into the Serial Monitor of the Arduino IDE. In the first part of the experiment:
\begin{itemize}
    \item \texttt{1} starts the background flow,
    \item \texttt{0} stops the background flow.
\end{itemize}

Pump control commands follow the pattern  
\texttt{Pump,Pump Duration,Duration}.  
For example, entering \texttt{1,3 30,50} runs Pump~1 for 30\,ms and Pump~3 for 50\,ms. single-pump commands such as \texttt{2 100} are also possible. Durations above 100\,ms should be avoided, as they flood the channel with dye and hinder measurement.

In the second part of the experiment, students may enter text messages, which are encoded and transmitted through the fluid channel. Messages should be kept short due to long transmission times.

An overview of some example commands can be found in Table~\ref{tab:commands}.

If uploading new code causes an error such as \texttt{Connecting\ldots}, pressing the \emph{Enable} button next to the USB-C port resolves the issue.

\begin{table}[h]
\centering
\caption{Example program commands}
\begin{tabular}{|l|l|l|l|}
\hline
\textbf{Program} & \textbf{Input} & \textbf{Output} & \textbf{Experiment Part} \\ 
\hline
\multirow{5}{*}{Pump Control} 
  & \texttt{1} & starts background flow & all parts \\
  & \texttt{0} & stops background flow & all parts \\
  & \texttt{1 30} & Pump~1 runs 30\,ms & Parts 1--2 \\
  & \texttt{1,3 30,50} & Pump~1 30\,ms, Pump~3 50\,ms & Parts 1--2 \\
  & \texttt{Hello World} & transmits message & Part 3 \\
\hline
\multirow{1}{*}{Sensor Control} 
  & \texttt{1} & calibrates sensor & after each re-upload \\                          
\hline
\end{tabular}
\label{tab:commands}
\end{table}

\subsection{Notes on the Photosensor Code}

The photosensor code allows both intensity curves and percentage intensity changes to be displayed during the first part of the experiment. In the second part, transmitted messages are detected and decoded, with the received message shown in the Serial Monitor. Before any measurements can be taken, the channel must be calibrated by placing a clean water-filled tube over the sensor and entering \texttt{1}.

\section{Experimental Procedure}
\label{sec:versuchsdurchfuehrung}

This section outlines the practical execution of the experiment, including the operational steps, safety considerations, and recommended handling procedures. A structured and careful workflow ensures reliable measurements and prevents damage to sensitive components such as pumps and sensors.

\subsection{General Procedure}

The experiment begins with the preparation of the fluidic system. All tubes must be filled correctly with distilled water prior to use to avoid air pockets that would impair pump performance. The background flow is then activated, and the dye channels are primed by injecting small amounts of dye into the system until the dye reaches the Y-connectors.

During operation, dye injections are performed according to the programmed pump commands. The photosensor continuously measures the optical signal and transfers the data to the connected computer for analysis. Care must be taken to use appropriate injection durations (typically no more than 100\,ms) to avoid overfilling the channel with dye, which would complicate signal detection.

Throughout the experiment, students must monitor the water reservoir to ensure that the background-flow pump does not run dry. Similarly, they must verify that the sensor calibration is performed whenever a new program is uploaded to the controller.

\subsection{Operational Safety and Handling}

To minimize contamination and ensure safe handling of the dyes, laboratory gloves and coats should be worn during setup and injection phases. Although less critical during computer-based tasks, protective clothing is highly recommended whenever liquids are handled.

Due to the physical characteristics of the system, pump and container placement must follow the guidelines outlined in Section~\ref{subsec:aufbau}:
\begin{itemize}
    \item Dye reservoirs should be placed lower than the main flow channel to prevent unintended backflow.
    \item The waste container should be positioned at the highest point to allow controlled drainage.
    \item Tubing should not be filled while pumps are electrically connected in order to avoid accidental overpressure and pump damage.
\end{itemize}

\subsection{Disassembly and Cleaning}

After completing the experiment, proper disassembly and cleaning of all components is essential to maintain equipment functionality and prevent cross-contamination in future runs.

Because small amounts of liquid may spill during disassembly, paper towels should be kept ready to absorb any leaked fluids. Any remaining dye in the syringes may be poured back into the corresponding dye bottles. Afterwards, the dye reservoirs used in the setup must be cleaned. For this purpose, they are filled with distilled water and emptied using the dye pumps while the background flow is running. This procedure ensures that the dye is flushed out through the system rather than spilling when disconnecting tubing from the pumps.

Once all tubes show no visible dye residue, the remaining water in the main reservoir may be emptied into the waste container. All tubes may then be disconnected from the connectors and pumps. Any residual water inside the separated tubing can be removed by pushing air through the tubes with an empty syringe.

Finally, all syringes, tubes, and containers used during the experiment should be laid out to dry thoroughly before storage.

\subsection{Measurement of Background Flow}\label{subsec:flow_calc}
\begin{table}[ht]
\centering
\caption{Measurement of the Flow Rate}
\begin{tabular}{|c|c|}
\hline
\textbf{Measurement} & \textbf{Time for Background Flow Volume (3 ml)} \\
\hline
1 & 35.13 s \\
\hline
2 & 34.84 s \\
\hline
3 & 35.9 s \\
\hline
\end{tabular}
\label{tab:messwerte}
\end{table}

From averaged background flow speed measurements, see examples in Table~\ref{tab:messwerte}, we can calculate the following characteristic values:

\vspace{0.5cm}

\begin{tabular}{@{} l l @{}}
\textbf{Background Flow:} &
\smash{\(Q_{\text{0}}=\frac{3\,\mathrm{ml}}{35.287\,\mathrm{s}}\)}
\(\approx 0.085\,\mathrm{ml/s}\) \\[1.5em]

\textbf{Average Flow Velocity:} & \(
v_{\text{avg}} = \frac{Q_{\text{0}}}{\pi \cdot r_c^2}
= \frac{0.085\,\mathrm{ml/s}}{\pi \cdot (0.65 \times 10^{-3}\,\mathrm{m})^2}
= \frac{0.085 \times 10^{-6}\, \mathrm{m}^3/\mathrm{s}}{\pi \cdot (0.65 \times 10^{-3}\,\mathrm{m})^2}
\approx 0.064\, \frac{\mathrm{m}}{\mathrm{s}}
\) \\[1.5em]

\textbf{Reynolds Number:} &
\(
Re = \frac{d_c \cdot v_{\text{avg}}}{v_{\text{water}}}
= \frac{1.3 \times 10^{-3}\,\mathrm{m} \times 0.064\, \frac{\mathrm{m}}{\mathrm{s}}}{1.01 \times 10^{-6}\, \frac{\mathrm{m}^2}{\mathrm{s}}}
\approx 82.38
\)
\\[1.5em]

\textbf{Péclet Number:} &  
\(
Pe = \frac{r_c \cdot v_{\text{avg}}}{D}
= \frac{0.65 \times 10^{-3}\,\mathrm{m} \times 0.064\, \frac{\mathrm{m}}{\mathrm{s}}}{2.299 \times 10^{-9}\, \frac{\mathrm{m}^2}{\mathrm{s}}}
\approx 18094.82
\)
\\
\end{tabular}

\vspace{0.5cm}

\textbf{Is the flow laminar or turbulent, and must diffusion be considered?}\\[0.5em]

The flow is laminar, because \(82.38 < 2100\). Diffusion does not need to be considered, since \(18094.82 >> 1\).

\subsection{Color Detection}

\begin{figure}[htb]
  \centering
  \includegraphics[width=1\textwidth]{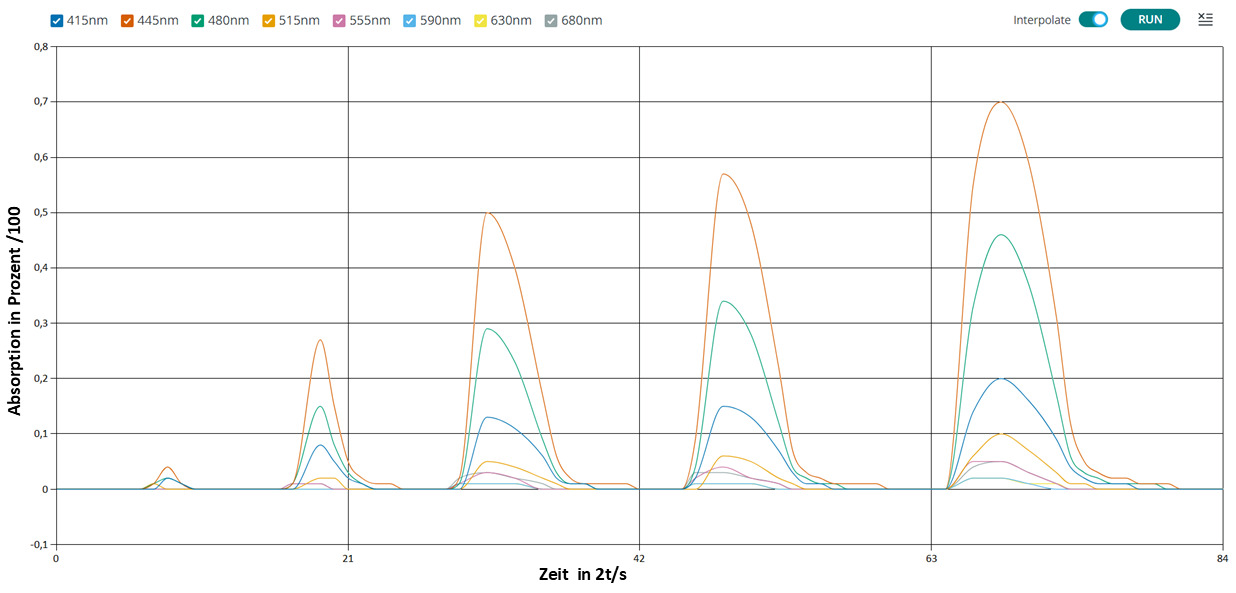} 
  \caption{Absorption of Yellow Dye for 15 ms, 30 ms, 45 ms, 60 ms, 100 ms}
  \label{fig:yellow_measurements}
\end{figure}

\vspace{0cm} 
\begin{table}[htb]
\centering
\caption{Absorption Values for Yellow Dye}
\begin{tabular}{|c|c|c|c|c|c|}
\hline
\textbf{Wavelength} & \textbf{15 ms} & \textbf{30 ms} & \textbf{45 ms} & \textbf{60 ms} & \textbf{100 ms} \\
\hline
415 nm & 0.02 & 0.08 & 0.13 & 0.15 & 0.20 \\ 
\hline
445 nm & 0.04 & 0.27 & 0.50 & 0.57 & 0.70 \\ 
\hline
480 nm & 0.02 & 0.15 & 0.29 & 0.34 & 0.46 \\ 
\hline
515 nm & 0.01 & 0.02 & 0.05 & 0.06 & 0.10 \\ 
\hline
555 nm & 0.00 & 0.00 & 0.01 & 0.01 & 0.02 \\ 
\hline
590 nm & 0.00 & 0.01 & 0.03 & 0.04 & 0.05 \\ 
\hline
630 nm & 0.00 & 0.00 & 0.01 & 0.01 & 0.02 \\ 
\hline
680 nm & 0.00 & 0.01 & 0.03 & 0.03 & 0.05 \\ 
\hline
\end{tabular}
\label{tab:yellow}
\end{table}

\begin{figure}[htb]
  \centering
  \includegraphics[width=1\textwidth]{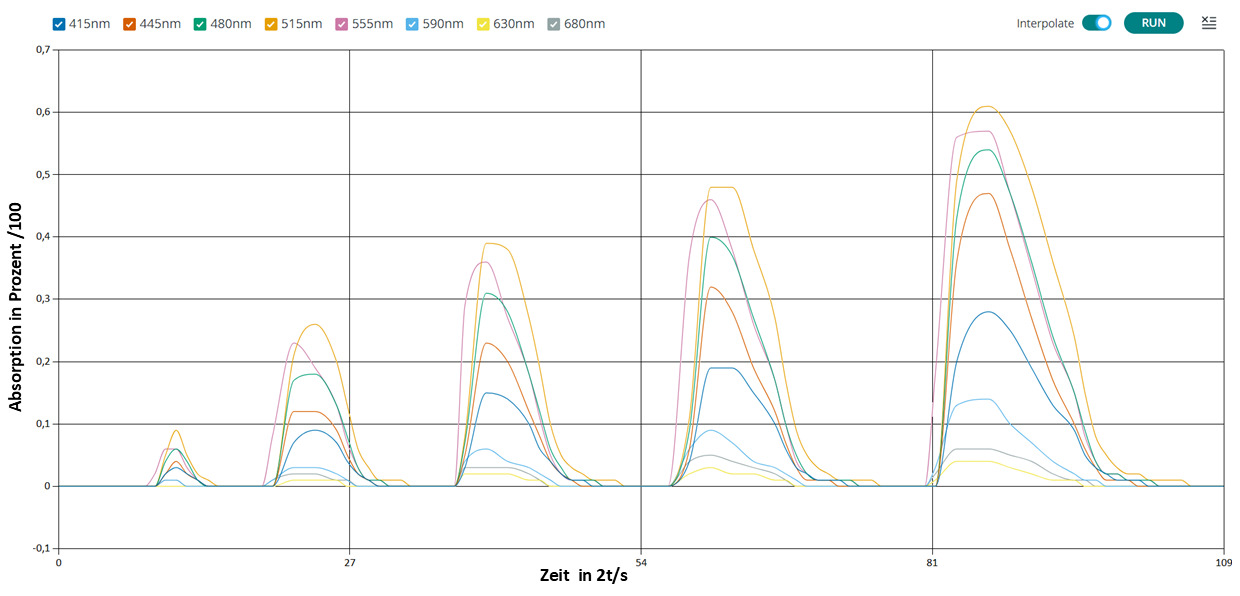} 
  \caption{Absorption of Magenta Dye for 15 ms, 30 ms, 45 ms, 60 ms, 100 ms}
  \label{fig:magenta_measurements}
\end{figure}

\vspace{0cm}
\begin{table}[htb]
\centering
\caption{Absorption Values for Magenta Dye}
\begin{tabular}{|c|c|c|c|c|c|}
\hline
\textbf{Wavelength} & \textbf{15 ms} & \textbf{30 ms} & \textbf{45 ms} & \textbf{60 ms} & \textbf{100 ms} \\
\hline
415 nm & 0.03 & 0.09 & 0.15 & 0.19 & 0.28 \\ 
\hline
445 nm & 0.04 & 0.12 & 0.23 & 0.32 & 0.47 \\ 
\hline
480 nm & 0.06 & 0.18 & 0.31 & 0.40 & 0.54 \\ 
\hline
515 nm & 0.09 & 0.26 & 0.39 & 0.48 & 0.61 \\ 
\hline
555 nm & 0.06 & 0.23 & 0.36 & 0.46 & 0.57 \\ 
\hline
590 nm & 0.01 & 0.03 & 0.06 & 0.09 & 0.14 \\ 
\hline
630 nm & 0.00 & 0.01 & 0.02 & 0.03 & 0.04 \\ 
\hline
680 nm & 0.01 & 0.02 & 0.03 & 0.05 & 0.06 \\ 
\hline
\end{tabular}
\label{tab:magenta}
\end{table}

\begin{figure}[htb]
  \centering
  \includegraphics[width=0.94\textwidth]{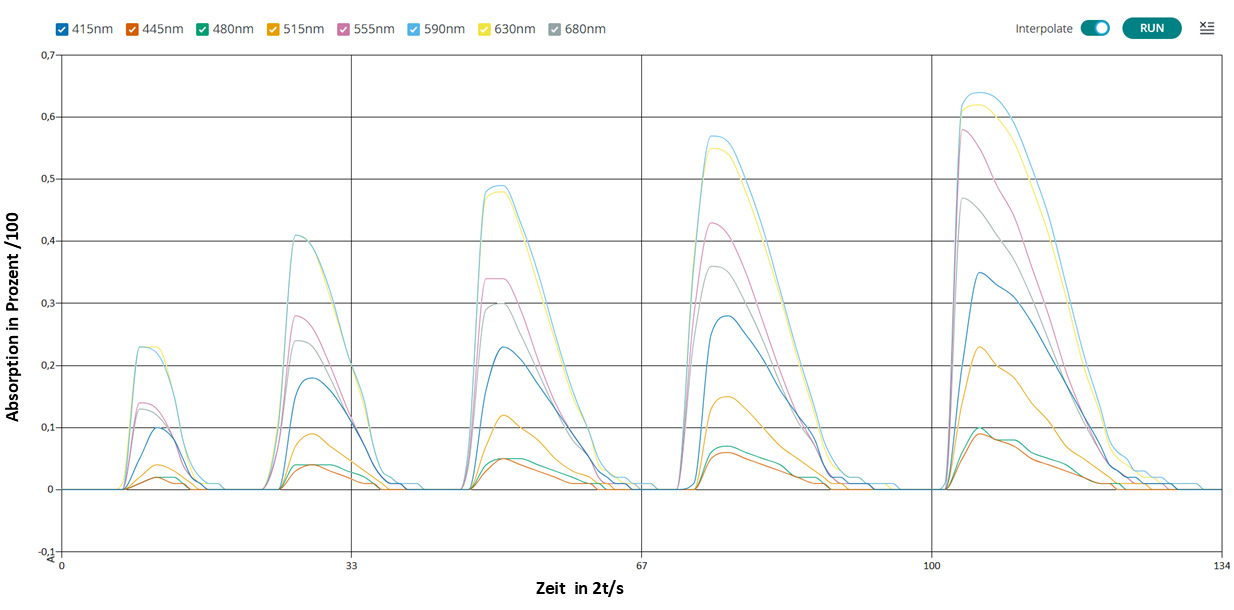} 
  \caption{Absorption of Cyan Dye for 15 ms, 30 ms, 45 ms, 60 ms, 100 ms}
  \label{fig:cyan_measurements}
\end{figure}

\begin{table}[htb]
\centering
\caption{Absorption Values for Cyan Dye}
\begin{tabular}{|c|c|c|c|c|c|}
\hline
\textbf{Wavelength} & \textbf{15 ms} & \textbf{30 ms} & \textbf{45 ms} & \textbf{60 ms} & \textbf{100 ms} \\
\hline
415 nm & 0.10 & 0.18 & 0.23 & 0.28 & 0.35 \\ 
\hline
445 nm & 0.02 & 0.04 & 0.05 & 0.06 & 0.09 \\ 
\hline
480 nm & 0.02 & 0.04 & 0.05 & 0.07 & 0.10 \\ 
\hline
515 nm & 0.04 & 0.09 & 0.12 & 0.15 & 0.23 \\ 
\hline
555 nm & 0.14 & 0.28 & 0.34 & 0.43 & 0.58 \\ 
\hline
590 nm & 0.23 & 0.41 & 0.49 & 0.57 & 0.64 \\ 
\hline
630 nm & 0.23 & 0.41 & 0.48 & 0.55 & 0.62 \\ 
\hline
680 nm & 0.13 & 0.24 & 0.30 & 0.36 & 0.47 \\ 
\hline
\end{tabular}
\label{tab:cyan}
\end{table}

The Tables~\ref{tab:yellow}, \ref{tab:magenta}, and~\ref{tab:cyan} represent the maximum deviations in wavelength detection between the dye-free channel with background flow and the channel after dye injection, of the yellow, magenta, and cyan dye, respectively.
The channel matrix is formed by column-wise concatenation of the values (Yellow, Magenta, Cyan) for an injection duration of $30$ ms.  The pseudoinverse computed for the given values is:

\[
\begin{bmatrix}
  0.674908 &  3.089582 &  0.990989 & -1.346267 & -1.164865 &  0.082153 &  0.212295 &  0.174566 \\
 -0.008261 & -0.632910 &  0.754025 &  2.297841 &  1.640989 & -0.600844 & -0.802874 & -0.392146 \\
  0.301127 &  0.017085 & -0.196371 & -0.336957 &  0.192583 &  0.948434 &  0.992527 &  0.55656936
\end{bmatrix}
\]

The values calculated in this experiment may vary slightly depending on the setup, but the general relationship between absorption values across wavelengths remains similar. Therefore, the pseudoinverse must be recalculated for each experiment.

\begin{figure}[h]
  \centering
  \includegraphics[width=1\textwidth]{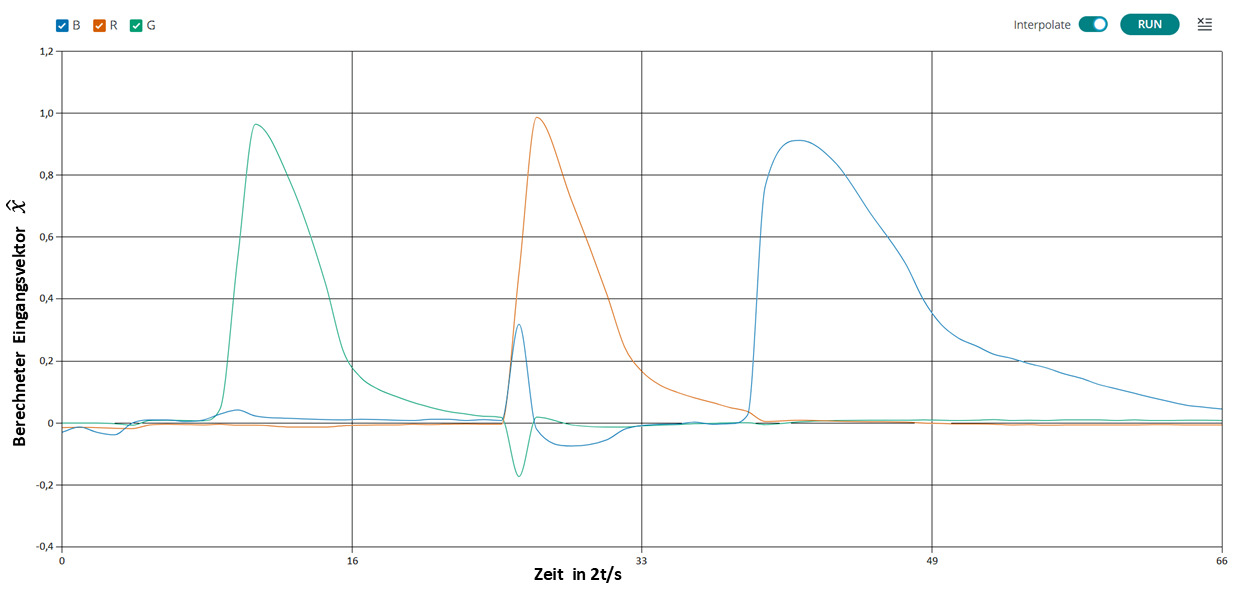} 
  \caption{Detected Color Values at the Receiver Using the Computed Pseudoinverse}
  \label{fig:pseudo_plot}
\end{figure}

The detected dye concentration after \ac{ZF} for three consecutive pulses of the three dyes is shown in Figure~\ref{fig:pseudo_plot}.

Example measurements of injections of the yellow, magenta, and cyan dye are plotted in Figures~\ref{fig:yellow_measurements}, \ref{fig:magenta_measurements}, and \ref{fig:cyan_measurements}. Related practical questions are discussed in the following.\\[0.5em]

\textbf{Why is the cyan curve wider than the yellow one?}\\[0.5em]

The cyan dye has more time to spread through the channel because it travels a longer distance before reaching the receiver. During the experiment, it can be clearly observed that the cyan dye spreads across a longer section of the channel. As a result, it is detectable for a longer period of time. This must be taken into account during data transmission by adjusting the thresholds or bit durations.

\vspace{2em}

\textbf{What value should the 30 ms injection achieve after applying the pseudoinverse, and why can this value not be measured exactly?}\\[0.5em]

After applying the pseudoinverse, the resulting value should ideally be 1, due to the \ac{ZF} method, which compares the similarity of the received signal to the calibration signal.  
In the experiment, noise and non-identical color behavior in the channel prevent an exact value of 1. With good calibration, the values fluctuate slightly around 1.

\vspace{2em}

\textbf{Which threshold value should be chosen for the colors and the start flag?}\\[0.5em]

The expected thresholds depend on the experiment’s measured values. In general, the thresholds for the colors must be set high enough so that the correct dye is always detected and intensity fluctuations do not cause false detections.  
Typical threshold values are around 0.6 to 0.8.  
For the start flag, a threshold must be chosen that is never reached by any of the dyes during standard operation.

\vspace{2em}

\textbf{Why do the values not increase linearly with pulse duration?}\\[0.5em]

The values do not increase linearly because the channel becomes saturated with dye. In the limiting case, the channel would contain only dye, and additional injections would only increase detection duration instead of intensity.  
The increase follows an exponential curve approaching an asymptotic saturation level.

\subsection{Data Transmission}

Table~\ref{tab:data_rate} shows an example configuration for the best data rate achieved during the lab.


\vspace{1.5em}
\begin{table}[H]
\centering
\caption{Best Data Rate Achieved}
\begin{tabular}{|l|l|}
\hline
\textbf{Parameter} & \textbf{Value}\\
\hline
Injection Duration – Dye: & 30 ms \\[1.5em]
Injection Duration – Flag: & 50 ms \\[1.5em]
Sequence Duration: & 6000 ms \\[1.5em]
Start Sequence – Transmitter: & 5000 ms \\[1.5em]
Start Sequence – Receiver: & 3000 ms \\[1.5em]
Threshold – Bit: & 0.6 \\[1.5em]
Threshold – Flag: & 1.4 \\[1.5em]
Data Rate: & \(0.5 \frac{bit}{s}\) \\[0.5em]
\hline
\end{tabular}
\label{tab:data_rate}
\end{table}

\vspace{2em}

This rate could be increased by using error-correcting mechanisms or by modifying the experimental setup, as dye behavior strongly depends on tube curvature. Measurement quality could also be improved if fixed, exact amounts of dye are injected each time.  
The reported data rate was obtained for a distance of 15 cm between transmitter and receiver.

\vspace{1.5em}

\textbf{Why can the data rate not be increased arbitrarily?}\\[0.5em]

A key limitation is the dye pumps: for injection durations below 15 ms, the pumps deliver too little dye to be reliably detected.  
Because the dye spreads along the tubing, thresholds must be high enough to prevent interference between consecutive sequences or detection failure due to very low concentrations.

\vspace{2em}

\textbf{What are potential ways to improve the achieved data rate?}\\[0.5em]

Possible answers include optimizing the thresholds so they reliably detect the dye while allowing shorter bit durations.  
Injection times may be reduced so that the dye spreads less and does not interfere with the next sequence.  
Error-correction mechanisms in software could detect and compensate individual bit errors.

\section{Conclusions}

In this work, we developed and evaluated a new hands-on laboratory experiment based on~\cite{wietfeld2025evaluation} that introduces undergraduate students to the fundamentals of \ac{MC}. The experiment provides a compact, reproducible, and low-cost testbed based on dye-based signaling in a fluidic channel, enabling students to explore core concepts such as flow-based transport, spectral detection, channel characterization, and \ac{ZF}--based color separation.

Pilot sessions with students demonstrated that the experiment can be carried out successfully within a single laboratory period and that the accompanying script offers a clear and comprehensible introduction to the topic. Students were able to implement the full communication chain from channel calibration and pseudoinverse computation to message transmission and decoding. They achieved stable data rates of up to 0.5~bit/s over a distance of 15~cm. Feedback indicated that the experiment provides an intuitive and engaging first encounter with \ac{MC}, even for participants with only basic prior knowledge of communication engineering.

Overall, the achieved objectives of this project are the following: a complete experiment, including a reproducible setup, detailed instructions, and an educational script, were designed, implemented, and validated. Beginning in the winter term 2025/2026, the experiment \emph{``A Testbed for Molecular Communication''} was utilized in the \textit{Communication Engineering} lab course, thereby modernizing the curriculum and exposing students to an emerging research area with relevance for future nanoscale and biomedical communication systems.

\section*{Acknowledgement}
C. Deppe, A. Gaedeken, W. Kellerer, A. Wietfeld, and Y. Zhao acknowledge the financial support of the Federal Ministry of Research, Technology and Space of Germany (BMFTR) within the programme “Souverän. Digital. Vernetzt.”, Joint Project 6G-life, under project identification numbers 16KISK002 and 16KISK263.
E. A. Jorswieck is supported by the Federal Ministry of Research, Technology and Space of Germany (BMFTR) within the programme “Souverän. Digital. Vernetzt.”, Joint Project 6G-Research and Innovation Cluster (6G-RIC), under Grant 16KISK031.
C. Deppe and Y. Zhao were also supported by the DFG within the project DE1915/2-1.

\bibliographystyle{IEEEtran}
\bibliography{bibliography}

\newpage

\appendix

\bigskip
  \begin{center}
\bfseries Material for Students\\[2em]
  
    \Large Communication Engineering Lab Course\\[2em]
    \huge\bfseries A Testbed for Molecular Communication\\[3em]
    \Large Institute for Communications Technology
  \end{center}

  \newpage

\tableofcontents
\newpage

 \section{Introduction}
            Molecular communication is a type of message transmission intended for use in areas where conventional optical, acoustic, or electromagnetic communication methods have proven impractical and inefficient \cite{MC_Nachrichtenidentifikation}. The possible applications are very diverse, including uses in biomedicine, health information, environmental monitoring and control, as well as in industrial settings \cite{Molecular_Communication_Networking}.\newline
Nature serves as a model for molecular communication. For example, the exchange of information between pests is based, among other things, on pheromones—special sexual attractants which, in the sense of bionics, are used in agriculture as pest control agents to prevent mating and thus protect crops \cite{Pheromone}. Plant protection is not the only area in which molecular communication is used; certain bacteria, for example, stimulate plant growth. Thus, by artificially producing these bacteria, it is possible to influence plant growth \cite{pflanzenwachstum}.
More general ecological applications are also conceivable, such as monitoring ecosystems and populations \cite{Pheromone}.
Monitoring and control are not limited to the agricultural context but are also applicable in industrial settings.\newline
It is envisioned to use molecular communication for monitoring chemical reactions or manufacturing processes at the nanometer scale. Applications such as the so-called ''e-nose''~for the food industry are promising, for example, to conduct quality control in breweries or other sectors of the food industry \cite{e-nose}.\newline
Molecular communication based on scents is also conceivable in everyday life, for example, by releasing molecules into the air to create an immersive experience in video games, allowing them to be perceived with all senses \cite{MC_Geruch}.
For industrial sectors that are difficult to access using communication via electromagnetic waves due to local conditions—such as mining with its tunnel systems—communication via airborne molecules appears to be a reliable alternative for message transmission \cite{MC_industrie_bergbau}. \newline
When used in the medical field, the possibilities are equally broad and promising. Since molecular communication is event-driven, it is suitable, among other things, for diagnostic applications such as detecting biomarkers, pH changes, and immune reactions. Such methods allow the detection of viral and bacterial diseases as well as tumors \cite{MC_Nachrichtenidentifikation}.
\newline To realize these applications at the level of molecular communication, the implementation of nanomachines on the order of 10–100 $,\mu\text{m}^2$ is necessary \cite{Nano_Networks}.
One of the main areas of current research in molecular communication focuses on attempts to realize disease detection and targeted drug delivery using such nanomachines operated through a lab-on-chip \cite{6G-life}. \newline
It is not only the treatment of infectious diseases that seems conceivable but also the treatment of degenerative diseases of synaptic connections, such as Alzheimer’s disease. For such purposes, brain–machine interfaces are being researched, which are intended to provide a system based on molecular communication capable of regulating, modifying, or mimicking synaptic pathways in the brain \cite{synapticCommunication}.\newline
For all this health data to be detectable from outside the body, another important research aspect of molecular communication is the development of a bio-cyber interface. This should, for example, be integrable into something like a wristwatch and provide an interface between molecular communication inside the body and conventional communication \cite{6G-life}. This is also an essential aspect for the implementation of the Internet of Things or the Internet of Bio-Nano Things, as the goal in the future 6G standard is to connect the human world with the digital and physical world. To achieve this, interfaces such as bio-cyber interfaces are required to enable communication between the human world and the external world \cite{6G_MC_möglichkeiten,internet_of_things_allgemein}. For message transmission within the human body, molecular communication is one of the most promising options for enabling future communication between multiple biological entities in the body that collect, process, and transmit data \cite{Internet_of_things}.
\newline For molecular communication to be incorporated into 6G standardization, significant progress is still required in information theory, nanotechnology, and biochemical cryptography \cite{MC_Nachrichtenidentifikation}.\newline
TU Braunschweig supports research in the field of molecular communication in connection with message identification, and the University of Erlangen has also introduced its first lectures on this topic. However, research in this field is not limited to communications engineering but must be understood as interdisciplinary, involving collaboration between biologists and scholars dealing with ethical aspects \cite{interdisziplinär}.\newline
As described, molecular communication can be used in many ways, but it is also highly complex, whether it operates via airborne molecules or molecules transported through liquids. To provide a simple introduction to this field and to better understand fundamental concepts, this work adapts the testbed developed at TU Munich~\cite{wietfeld2025evaluation} and establishes molecular communication using inks in a liquid medium.
            \section{Fundamentals} 
            \section{Types of Molecular Communication}
                For the implementation of molecular communication, there are various approaches. These differ mainly in that, in some models, the molecules are released directly into the corresponding medium, while in others the molecules are transported from the transmitter to the receiver in the form of a ''motor''.


\subsection{Diffusion}
Communication via diffusion is a principle that the human body naturally uses to convey messages. Here, molecules or signaling substances are released by a transmitter into a medium such as blood, where they are then absorbed by a receiver \cite{diffusion}.

\subsection{Gap junction channel}
In nature, there is also the so-called gap junction channel, which can be used in molecular communication. In this case, molecules are propagated between individual cells through natural channels that connect the cytosol of adjacent cells.

\subsection{Molecular motors}
The approach using molecular motors follows the idea that one or several molecules are contained within a motor that moves along a kind of track. This movement works by forming and releasing bonds with the track \cite{mallik2004motors}.

\subsection{Bacterial transport}
In nature, bacteria move using flagella—protein filaments that can rotate like a propeller. This type of movement can be used for molecular communication. In this approach, the molecules are enclosed in an artificial bacterium that is guided to the transmitter using a nutrient source \cite{bakterien_transport}.

\subsection{Comparison of communication variants}
Each of these variants of communication using molecules has its own advantages and disadvantages. The question then arises: which of these variants is suitable for an application with nanorobots on the scale of a blood platelet, so that such systems could later be used, for example, in the human body? Essential factors to consider include the limited surface area available for implementation as well as the low energy resources \cite{gohari2016information}.\
The main disadvantages of molecular motors, gap junctions, and artificial bacteria are, on the one hand, the additional infrastructure required—which is not practical—and, on the other hand, the high energy consumption of molecular motors.\
Thus, the most feasible form is communication via diffusion, since the molecules only need to be introduced into the corresponding medium and no infrastructure is required. However, the transport of molecules from the transmitter to the receiver is completely uncontrolled, meaning the molecules have different propagation times from transmitter to receiver, which can lead to channel distortion and loss of information \cite{nakano2013molecular}.
            \section{Diffusion-Based Transmission Channel}
                By choosing diffusion-based molecular communication, certain specific characteristics arise for the design of the communication channel, which must be addressed. One of these is the molecular motion of particles, which is described by Brownian molecular motion, resulting in the distribution of molecules being statistically normally distributed \cite{gohari2016information}.

\subsection{Statistical distribution in the channel}
In a medium, the propagation of molecules occurs according to a concentration gradient, which can be described on a macroscopic level using Fick's laws.\\
The first Fick's law in the one-dimensional case is described by the particle flux density \(J\), the diffusion coefficient \(D\), and the concentration gradient \( \frac{\partial \rho}{\partial x} \):

\begin{equation}
J(x,t) = -D \frac{\partial \rho(x,t)}{\partial x} 
\end{equation}
With the mass conservation law, taking into account the density of molecule production \(c(x,t)\) at a point \(x\).

\begin{equation}
\frac{\partial \rho(x,t)}{\partial t} = - \frac{\partial J(x,t)}{\partial x}+ c(x,t)
\end{equation}
By inserting Eq. (2.2) into Eq. (2.1), we obtain the second Fick's law, extended by the density of molecule production.

\begin{equation}
\frac{\partial \rho(x,t)}{\partial t} = D \frac{\partial^2 \rho(x,t)}{\partial x^2}+ c(x,t)
\end{equation}
To solve this differential equation, we assume no boundary conditions that lead to reflection or absorption at the boundaries. Thus, solving this differential equation on the interval \(I =  (-\infty, \infty) \) with a transmitter at \mbox{$x=0$}, which releases a molecule at time \(t=0\), i.e., with the molecule production rate \(c(x,t)=\delta(x=0)\delta(t=0)\), leads to

\begin{equation}
\rho(x,t) = \frac{\mathbf{1}[t>0]}{(4\pi Dt)^{0,5}}e^{-\frac{x^2}{4Dt}}
\end{equation}

For a fixed time \(t\), it becomes apparent that the solution has the form of a Gaussian distribution over \(x\). It should be noted here that due to external influences such as chemical reactions or absorption of molecules, an analytical solution does not necessarily exist.\\
On a microscopic level, each molecule is subject to Brownian motion, which can be represented as a random walk process. In the one-dimensional case, this implies that each molecular movement is independent of the previous movement and can therefore be considered as a separate normally distributed random variable. Since the sum of normally distributed random variables is again normally distributed, it follows for \( 0 \leq t_1 < t_2 \) and for a molecule released at time \(t=0\) at the origin that its displacement follows \(\mathcal{N}(0, 2D(t_2-t_1))\). In summary, for a molecule found at time \(t\) at \(\rho(x,t)\), the variance is \(2Dt\), and \(\rho(x,t)\) has the same form as in Eq. 2.4. This is also supported by Einstein's 1905 publication on the relationship between Brownian motion and Fick's laws \cite{gohari2016information,ficksches_Gesetz}.

\subsection{Receiver}
There are two approaches to designing receivers. In one approach, molecule concentrations are measured, while the molecules are not absorbed from the medium. In the other approach, the molecules are absorbed from the medium or react with it.

\subsubsection{Non-reactive receivers}
Non-reactive receivers include the sampling receiver and the transparent receiver. The sampling receiver measures the concentration at a specific spatial point; similarly, the transparent receiver measures the concentration within the volume of a sphere \(V_R\). \newline However, it is assumed that these two receivers perfectly detect the number of molecules. The essential difference is that the sampling receiver measures the mean value \(\rho(x,t)\), while the transparent receiver measures the mean value \(\rho(x,t)\) and the variance \(\rho(x,t)/V_R\). For non-reactive receivers, the molecules are not absorbed, and the molecule concentration remains unchanged, so Fick's laws apply without boundary conditions \cite{gohari2016information}.

\subsubsection{Reactive receivers}
Reactive receivers include the absorbing receiver, which absorbs incoming molecules and thus changes the molecule concentration. Additionally, the ligand or reactive receiver functions analogously to cellular receptors. Here, molecules react with receptors on the surface of the receiver. These receptors can exist in the states '' bound''\\ or '' unbound''~, depending on whether the receptor has reacted with a molecule or not. Molecules cannot bind to receptors that are already bound, and the dissociation of molecule–receptor bindings occurs according to a temporal random process. The number of bound receptors can be modeled using a memoryless binomial distribution binomial \((k,p)\), with \(k\) as the number of receptors and \(p\) as the binding probability.

For the reasons mentioned, signals at the receiver can be modeled as \(Z=\alpha X+N\), where \(Z\) is the measured output at the receiver, \(\alpha\) is an amplification factor, \(X\) is the molecule concentration, and \(N\) is Gaussian noise.
\newpage
For Fick's laws to apply to both non-reactive and reactive receivers, special boundary conditions must be considered.
The surface of the absorbing receiver is modeled as a zero boundary, whereas for the ligand receiver it is modeled as a partially absorbing surface with an active molecule source 
 \cite{gohari2016information}.

            \section{Molecules}
                The design of molecular communication also requires the specification of the molecules in order to perform measurements. In other experiments, several approaches have already been tested. A distinction is made between receivers that communicate using a single molecule type according to Single-Molecule Communication (SIMO) and receivers based on the principle of Multi-Molecule Communication (MUMO), which use several molecule types for communication \cite{wietfeld2025evaluation}.

\subsection{Single-Molecule Communication}
One approach in SIMO is to communicate through the use of acids and bases. In these experiments, acids/bases are released and transported to the receiver—a pH probe—by several peristaltic pumps. Using recurrent neural detectors, data rates of up to 4 \(\frac{bit}{s}\) can be achieved \cite{MC_Base_Acid}.\\
Furthermore, switchable fluorescent proteins are used, whose main advantage lies in the repeated use of the same protein for different transmission cycles. For this type of experiment, two LED arrays at the transmitter are required for sending and erasing, which activate and deactivate the proteins. At the receiver, spectrometers are used to measure the fluorescence, enabling data rates of 0.1 \(\frac{bit}{s}\) \cite{floureszierendeProteine}.\\
So far, these two experimental approaches have only been conducted over communication distances of a few centimeters. For communication over longer distances, special magnetic nanoparticles were released by peristaltic pumps and received by a non-invasive susceptometer. Over short distances of 5 cm, a data rate of 6 \(\frac{bit}{s}\) is achieved, and messages can still be decoded over 40 cm \cite{MC_magneticParticels}.

\subsection{Multi-Molecule Communication}
For MUMO, several molecules are used for data transmission; for this, different colors can be employed, for example~\cite{wietfeld2025evaluation}.
In the subsequent experimental part of this exercise, this type of MUMO communication is to be carried out using different colors. For this communication, a linear estimator is used for decoding, which determines the time-dependent color intensities from the raw data of the spectral measurements of the photosensor for each color used. The data rates achieved with this method exceed 3 \(\frac{bit}{s}\) over a distance of more than 20 cm \cite{wietfeld2025evaluation}.

        \section{Fundamentals of the Laboratory Experiment}
            \section{Channel}
                For an optimal execution of the experiment, knowledge about the channel present in the setup—in this case, the water-filled tube—and in particular the behavior of the ink, is essential.
The movement of the ink molecules through the channel is based on advection, meaning they are transported from the transmitter to the receiver by the background flow of the water \cite{wietfeld2025evaluation}.
For flows in liquid media, it is also important to consider the type of flow as well as the diffusion of the molecules in order to better predict the behavior of the molecules in the medium.

\subsection{Types of flow}
In general, when dealing with the movement of liquid and gaseous media, a distinction is made between laminar and turbulent flows. Turbulent flows are flows in which vortices occur, whereas laminar flows are vortex-free. The classification of flows is based on the so-called Reynolds number \([Re]\)

\begin{equation}
 Re = \frac{d_c*v_{avg}}{v}
\end{equation}

with \(d_c\) denoting the diameter of the channel, \(v\) the kinematic viscosity, which for water is \(1.01*10^{-6} \frac{m^2}{s}\), and \(v_{avg}= \frac{Q_0}{\pi*r_c^2}\) the average water velocity, where \(Q_0\) corresponds to the background flow rate.\\
The transition from laminar to turbulent flow occurs at a Reynolds number of \(Re =2100\), with flows at \(Re <2100\) considered laminar. The experiment is based on a laminar flow. To determine the flow velocity in circular tubes as a function of the spatial point, the Poiseuille flow profile can be used:

\begin{equation}
    v(\rho) = v_{\text{max}} \left(1 - \frac{\rho^2}{r_c^2}\right), \quad \rho \in [0, r_c]
\end{equation}

The maximum flow velocity required in the formula is given by \(v_{max}=2*v_{avg}\) \cite{wietfeld2025evaluation,white2016fluid}.

\subsection{Diffusion}
The influence of diffusion on the transport of the ink can be determined using the 
P\'eclet number \([Pe]\). It is calculated as

\begin{equation}
    Pe =\frac{r_c*v_{avg}}{D}
\end{equation}

For the experiment, using the diffusion coefficient \([D]\) of water with \(D \underset{\sim}{<} 2.299*10^{-9} \frac{m^2}{s}\), it can be determined that the resulting P\'eclet number is significantly greater than one, meaning that the background flow dominates over diffusion. Therefore, diffusion can be neglected in the later execution of the experiment \cite{wietfeld2025evaluation,white2016fluid}.

\section{Determination of colors in the channel}

During the execution of the laboratory experiment, three different ink colors are injected in a controlled manner into the channel at the transmitter, and they must be detected separately at the receiver. For this purpose, ‘zeroforcing’ is applied in the experiment.\newline
\newline
In zeroforcing, it is assumed in mobile communication that a certain number of transmit antennas \(M_t\) and receive antennas \(M_r\) are present. For the laboratory experiment, zeroforcing can be applied analogously; the different colors injected into the channel represent \(M_t\), and the wavelengths of the different ink colors detectable at the receiver correspond to \(M_r\).\newline
For transmission, a channel matrix \(H\) can therefore be defined with \( H \in \mathbb{C}^{M_r \times M_t} \). To identify the different ink colors using zeroforcing, a MIMO equalization matrix \( A \in \mathbb{C}^{M_t \times M_r} \) is required. This matrix is used to assign the wavelengths to the individual colors by multiplying it with the channel matrix and the input vector.\newline
The result at the receiver is expressed as \(y=H*x+N\), where \(N\) denotes the noise variable. For the detection of colors using zeroforcing, this leads to \(\tilde{x} = A y\) with \(\tilde{x}(y) \in \mathbb{C}^{M_t}\). The estimated input vector is then mapped to

\begin{equation}
\hat{x}(y)_i = \arg\min_{s \in X} \left| \tilde{x}(y)_i - s \right| \quad \text{for all } i
\end{equation}

resulting in the input vector \(\hat{x}\), which is most similar to the computed one.\newline
To apply these theoretical foundations to the experiment, the channel matrix must be determined for all colors together. The simplest method is to examine the impulse response at the receiver for each color. This yields \( H \in \mathbb{C}^{8 \times 3} \), for which \(A\) must be determined; for square matrices, \(A\) can be obtained by direct inversion, whereas for non-square matrices, \(A\) must be computed using the pseudoinverse:

\begin{equation}
A=H^\dagger = (H^H H)^{-1} H^H
\end{equation}

Note that \(H^H\) is the Hermitian transpose of \(H\), and \(H^\dagger\) is the Moore–Penrose inverse. Thus,

\begin{equation}
\tilde{x} = A y
\end{equation}

In principle, \(y\) also contains a noise component, which is amplified by zeroforcing. However, this noise component can be neglected in the experiment because the signal strength—corresponding to the intensity of the colors—is significantly greater than the background noise.
Finally, the individual colors in \(\tilde{x}\) are detected using threshold values \cite{goldsmith2005wireless}.

               \section{Experiment}
            \section{Experimental Setup}
                In this experiment, water is pumped from a water reservoir through a tube using a pump, thereby generating a background flow. Into this flow, ink in the colors cyan, magenta, and yellow can be injected using three additional pumps and injection needles embedded in Y-connectors with epoxy resin. Built-in check valves prevent water–ink mixtures from flowing back into the ink supply tubes. The mixture of water and ink then flows past a photosensor, which is capable of detecting individual wavelengths and assigning them to the corresponding colors using programmed code. Finally, the water–ink mixture flows into a waste container for later disposal.

\begin{figure}[H]
    \centering
    \includegraphics[width=1\textwidth]{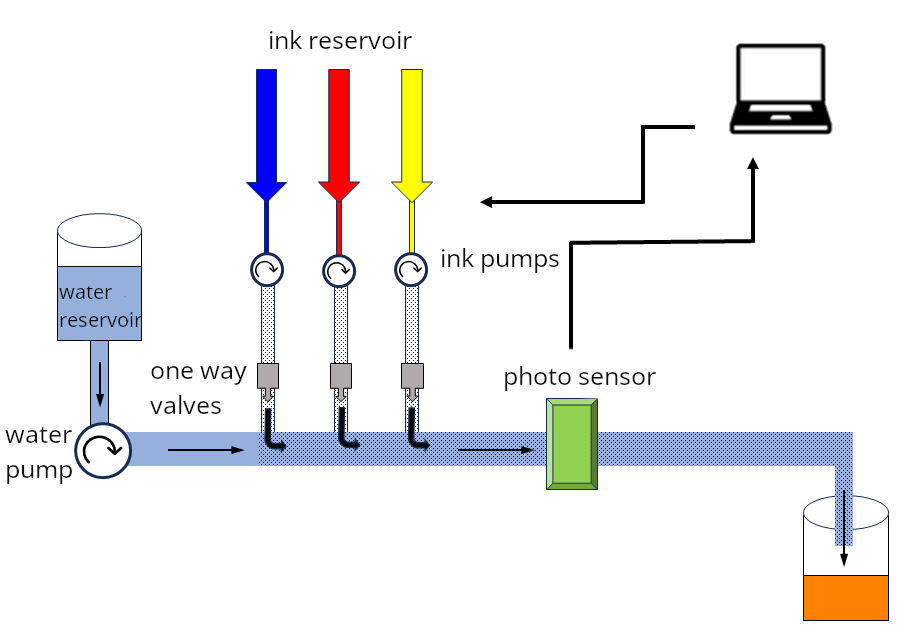}
    \caption{Sketch of the experimental setup}
    \label{apxfig:meinbild1}
\end{figure}

\newpage

\subsection{Pumps}

\begin{figure}[htbp]
    \centering
    \includegraphics[width=0.4\textwidth]{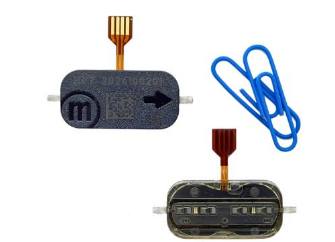}
    \caption{BP7–Tubing Pump}
    \label{apxfig:meinbild2}
\end{figure}

Piezoelectric membrane pumps (BP7–Tubing Pumps) are used for pumping. Piezoelectric means that the materials deform when an electrical voltage is applied; this principle is used in such pumps to transport liquids or gases with a pressure of up to 500 mbar. For safe operation, it is important that no voltage outside the range of 0–250 V is applied. For transporting liquids, only a modulation frequency of 100 Hz is required; higher frequencies are used exclusively for gases.\newline
It must be strongly emphasized that the piezoelectric pumps are very pressure-sensitive. The breakthrough pressure is 1.5 bar; if this pressure is exceeded, the pump will be damaged. In this experiment, syringes are used to store the inks and to initially fill the tubes. Due to the breakthrough pressure of the sensitive pumps, it is essential to avoid filling the tubes with liquid while the pumps are connected in order to prevent damage \cite{bartels2024}.\newline
The pumps are controlled via an ESP32 microcontroller installed on an mp-Multiboard2.

\subsection{Photosensor}

An Adafruit AS7341 10-Channel Light Sensor is used as the photosensor. It can measure nine different wavelengths, eight of them in the visible range (415 nm, 445 nm, 480 nm, 515 nm, 555 nm, 590 nm, 630 nm, 680 nm) and one in the near-infrared range. The sensor also includes a channel for measuring the total incoming light intensity.
The sensor is connected to an Arduino Micro via an \(I^2C\) interface, which reads the values and forwards them to the connected computer \cite{adafruit_as7341}.

\begin{figure}[H]
    \centering
    \includegraphics[width=1\textwidth]{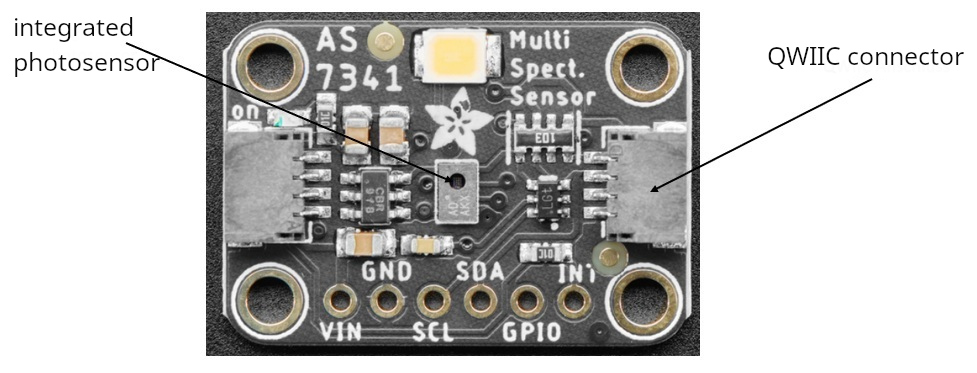}
    \caption{Adafruit AS7341 10-Channel Light Sensor}
    \label{apxfig:meinbild3}
\end{figure}

To avoid disturbances caused by ambient light, the photodiode is placed inside a 3D-printed enclosure. To ensure constant illumination of the sensor, a white LED—controlled by the Arduino Micro—is mounted above the photodiode. To measure the changes caused by the injected inks, the tube carrying the water–ink mixture is guided between the LED and the photodiode.
During the experiment, only the wavelengths in the visible range are read out.

\newpage

\subsection{Check valve}

\begin{figure}[H]
    \centering
    \includegraphics[width=0.3\textwidth]{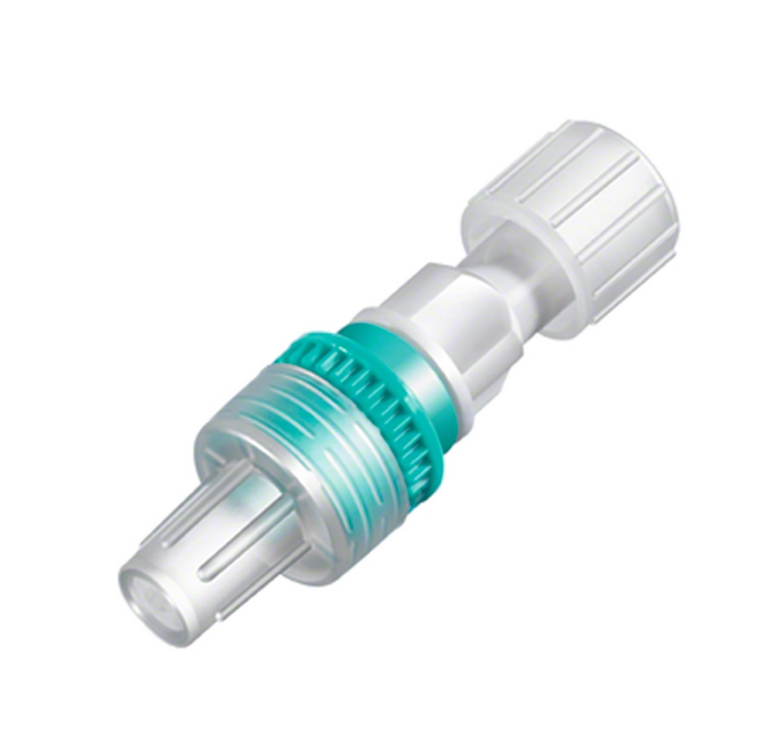}
    \caption{Check valve Infuvalve B. Braun \cite{infuvalve_6bar}}
    \label{apxfig:meinbild4}
\end{figure}

To prevent backflow of water into the ink and thus prevent dilution, check valves are used. These check valves are permeable only in one flow direction—from the white to the green side—and open at a pressure of 20 mbar. The pumps can generate a pressure of up to 500 mbar, ensuring that the check valves will open. Up to a backpressure of 2 bar, the valves reliably seal the channel. Neither the ink pumps nor the general water pressure will reach this level, ensuring that water cannot unintentionally enter the ink supply.\newline
To avoid damage to the check valves, no pressure exceeding 6 bar must be applied in the reverse direction. Therefore, when constructing the experiment, care must be taken to install the valves in the correct orientation to avoid damage during filling—even though manually creating such high pressures with syringes is unlikely  
\cite{infuvalve_2bar,infuvalve_6bar}.

\newpage

\subsection{Materials}
The following materials are required for the experiment:

\begin{table}[H]
\centering
\resizebox{\textwidth}{!}{
\begin{tabular}{|l|l|l|}
\hline
\textbf{Category}       & \textbf{Components} & \textbf{Manufacturer number} \\ 
\hline
\multirow{9}{*}{Electronic components} 
 & Arduino Micro &\\
 & Jumper cables &\\
 & Adafruit AS7341 10-Channel Sensor &4698\\
 & mp-Multiboard2 & Bartels Mikrotechnik: BM-S-0008\\
 & Micro-USB cable & Bartels Mikrotechnik: BM-S-0008\\
 & Bartels Pump | BP7-Tubing & Bartels Mikrotechnik: BM-S-0008\\
 & mp-Highdriver4 & Bartels Mikrotechnik: BM-E-0003\\
 & LED white & C513AWSNCX0Z0342\\
 & Resistor 4.7 k\(\Omega\) & MFR100FTE734K7\\
\hline
\multirow{7}{*}{Experiment components} 
 & 30G x 1/2" needles & PT9969 \\
 & 3 ml syringe &\\
 & TYGON LMT-55 tube AD[30mm] ID & Techlab: ISM SC0039T\\
 & Luer Lock adapter m (male) & Techlab: UP P-850 \\
 & Luer Lock adapter f (female) & Techlab: UP P-857 \\
 & Y-connector 1/16" &\\
 & Check valve Infuvalve\textregistered & B. Braun PZN: 02232430\\
\hline                            
\multirow{8}{*}{Other materials} 
 & Cyan ink & Conrad: 2233365-VQ\\
 & Yellow ink & Conrad: 2233367-VQ\\
 & Magenta ink & Conrad: 2233366-VQ\\
 & Distilled water &\\
 & Waste container &\\
 & 3D-printed water container &\\
 & Gloves &\\
 & Lab coat &\\
\hline
\end{tabular}
}
\caption{Overview of materials used}
\label{apxtab:ueberkategorien2}
\end{table}

\newpage

\subsection{Assembly instructions}
To avoid damaging any components during assembly, strict adherence to the assembly instructions is essential.\newline
First, the background flow must be set up. Before attaching any tube, each one must be filled with distilled water, as the pumps cannot operate properly if a water–air mixture is present in the tubes. To fill the tubes, a clean syringe with a Luer Lock (female) adapter is used to push water through the tubes until a continuous water column is established. These filled tubes are then connected to the water reservoir, the Luer Lock adapter (female), and the pump for the background flow. The flow direction is indicated by an arrow on the pump.\newline
From this pump, a tube must be connected to the Y-connector of the cyan ink; from this Y-connector, two additional tubes lead via two further Y-connectors to the magenta and yellow inks. From the final Y-connector (yellow ink), a tube is routed through the photosensor housing to the waste container.

\vspace{2em}
\begin{figure}[H]
    \centering
    \includegraphics[width=0.8\textwidth]{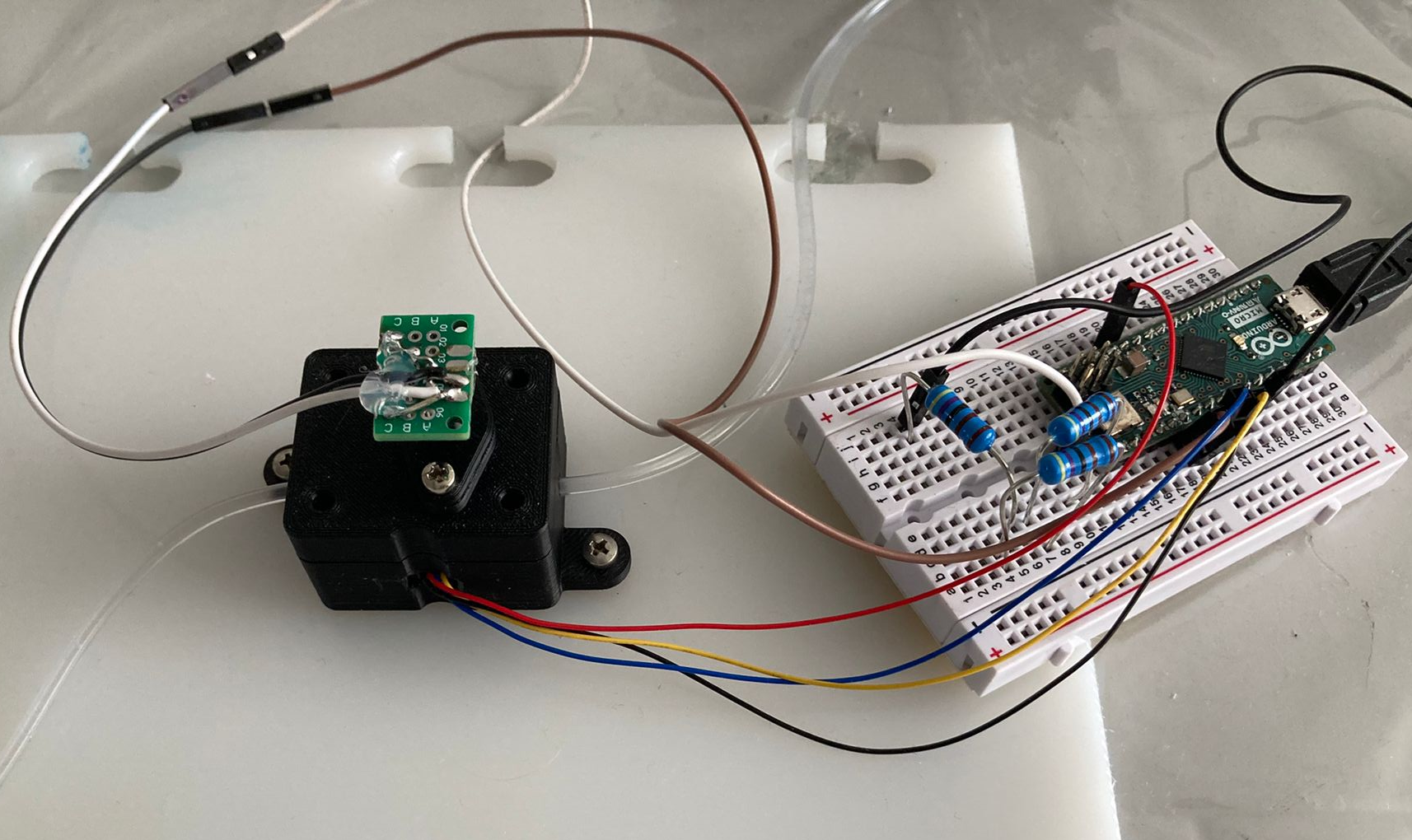}
    \caption{Experimental setup at the photosensor}
    \label{apxfig:meine_grafik1}
\end{figure}

\newpage

To connect the ink reservoirs, it is advisable to also fill the tubes leading away from the ink pumps with water and connect them to both the pump outlet and the white end of the check valve. The injection needle glued into the Y-connector is screwed into the green end of the check valve. To connect the valve, a Luer Lock (male) adapter must be attached. The tubes leading from the inks to the pump should first be filled with water, as filling directly with ink risks spilling ink onto the laboratory table.

\vspace{2em}
\begin{figure}[H]
    \centering
    \includegraphics[width=0.8\textwidth]{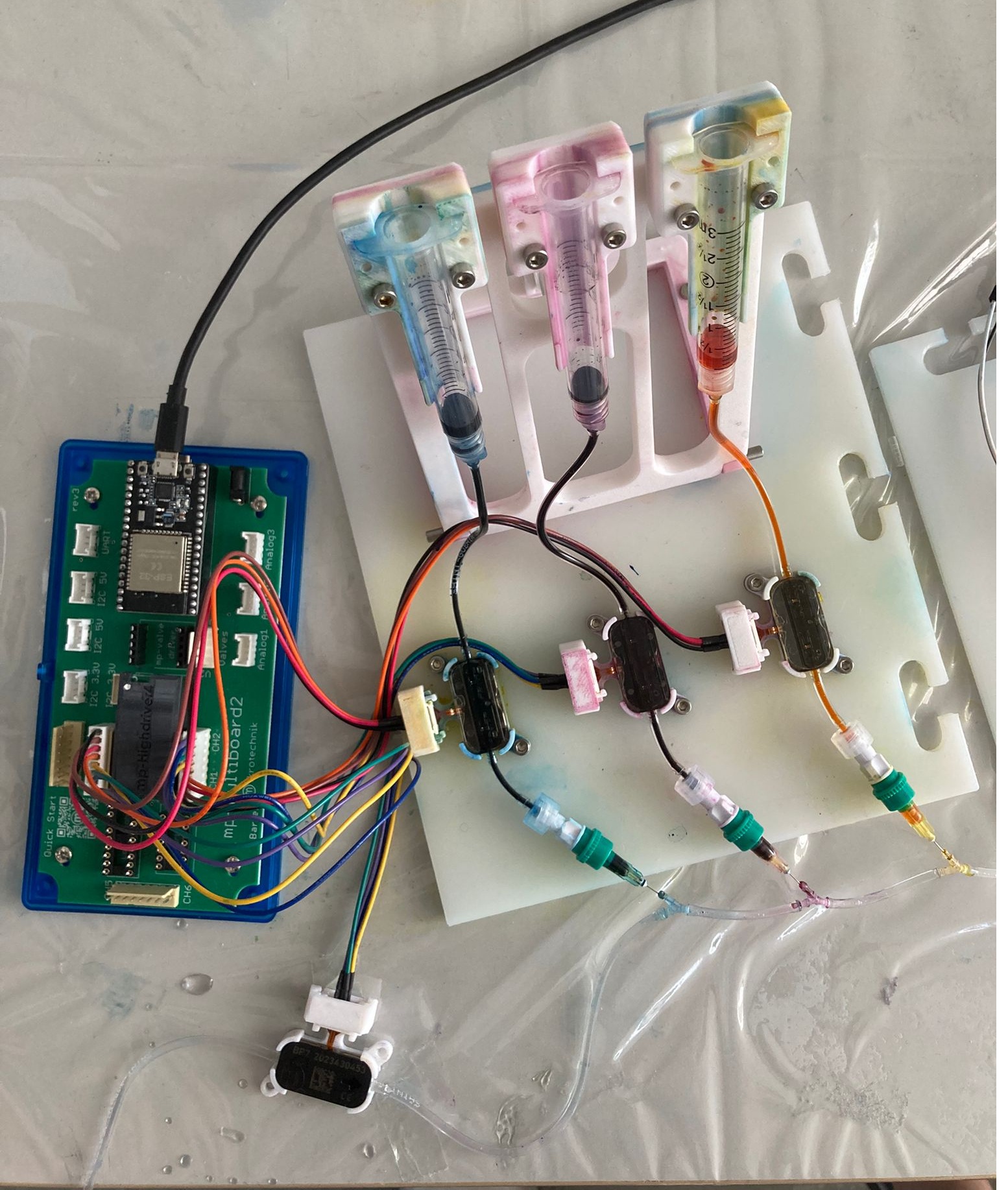}
    \caption{Ink injection setup}
    \label{apxfig:meine_grafik2}
\end{figure}

\newpage

In the final assembly step, the inks must be filled into their respective containers, and both the ESP32 and the Arduino Micro must be connected to the computer. Before starting the experiment, the water in the tubes from the ink reservoirs to the Y-connector must be pumped out. For this, the pump control program on the Multiboard should be used, and the inks should be pumped to the Y-connectors by manually entering the pump numbers and pump durations. If ink enters the tube of the background flow, it can be flushed out using the background flow pump.

\begin{figure}[H]
    \centering
    \includegraphics[width=0.8\textwidth]{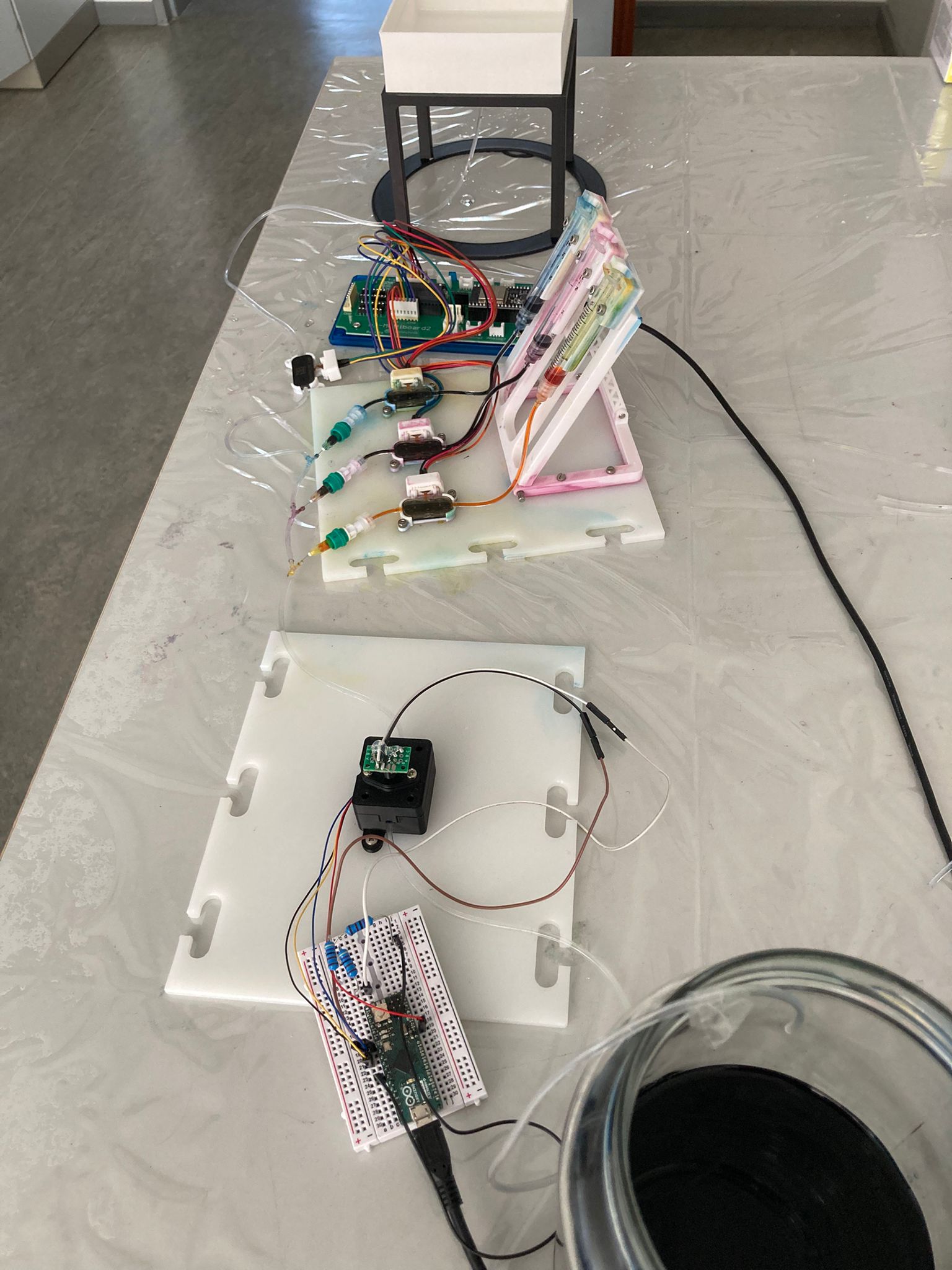}
    \caption{Complete experimental setup}
    \label{apxfig:meine_grafik3}
\end{figure}

\subsection{Notes on the experimental setup}
For the setup, it is recommended to use the provided gloves and lab coat to prevent ink from coming into contact with skin or clothing.\newline
The positioning of the containers is highly relevant, as the pumps are passively permeable in both directions when turned off. Although the check valves prevent water from entering the ink, they only partially prevent ink from flowing into the background-flow tube. Therefore, the ink injection points should be placed lower than the water level at the injection point into the tube or the outflow of wastewater.\newline
The reason ink flows out when the ink containers are positioned higher is that the liquid in a tube will settle to the same height on both sides due to the pressure of the water column. Thus, the wastewater container should be placed at the highest position, and the ink containers at the lowest, to prevent unintended ink flow that could lead to incorrect color detection at the photosensor.

\subsection{Notes on pump control}
The ink pumps are numbered from right to left as pumps 1 to 3 in the program code. For correct control of the pumps, they must be connected to the channel with the corresponding number on the mp-Multiboard2, and the background flow must be connected to channel 4. The pumps are controlled via the Serial Monitor in the Arduino IDE. In the first part of the experiment, the commands '' 1''\ (start background flow) and '' 0''\ (stop background flow) are available.\newline
Using the command format '' PumpNo,PumpNo Duration,Duration'', pumps can be activated. For example, ''1,2 100,100''\ runs pump 1 for 100 ms and pump 2 for 100 ms. A single-pump command may look like ''2 100''.~It is important to note that durations are given in milliseconds, and durations longer than 100 ms should be avoided, as they may flood the channel with ink and hinder measurement.

In the second part of the experiment, strings of letters can be entered into the Serial Monitor for message transmission. Since transmission takes a long time, messages should be kept short.\newline
In some cases, an error may occur when uploading code to the ESP—for example, if the ESP is still executing the previous code. If the terminal shows '' Connecting'', the problem can be solved by pressing the Enable button to the right of the USB-C port.

\begin{table}[H]
\centering
\resizebox{\textwidth}{!}{
\begin{tabular}{|l|l|l|l|}
\hline
\textbf{Program} & \textbf{Input} & \textbf{Output} & \textbf{Experiment part} \\ 
\hline
\multirow{5}{*}{Pump control} 
  & 1 & starts background flow & all parts \\
  & 0 & stops background flow & all parts \\
  & 1 30 & runs pump 1 for 30 ms & parts 1 and 2 \\
  & 1,3 30,50 & runs pump 1 for 30 ms and pump 3 for 50 ms & parts 1 and 2 \\
  & Hello World & transmits the message “Hello World” & part 3 \\
\hline
\multirow{1}{*}{Photosensor control} 
  & 1 & calibrates the channel & all parts (after re-upload)\\                          
\hline
\end{tabular}
}
\caption{Examples of possible program commands}
\label{apxtab:ueberkategorien1}
\end{table}

\subsection{Notes on the photosensor program code}
The program code for the photosensor enables both intensity curves and percentage intensity changes to be graphically displayed in the first part of the experiment. In the second part, the transmitted messages are detected and decoded, and the received message is displayed in the Serial Monitor. To perform measurements, when the channel is clean—meaning only water is present above the photosensor—a '' 1''~ must be entered.

            \section{Dismantling the Setup and Cleaning}
                For the correct handling of the laboratory materials, proper dismantling of the experiment and cleaning of the materials are necessary. A few points must be taken into account.\newline
Since liquid can always leak during dismantling, paper towels should be kept ready to absorb spilled liquids. Any remaining unused ink in the syringes can be poured back into the corresponding colored ink bottles. Afterwards, the ink reservoirs must be cleaned: for this purpose, the ink reservoirs of the experimental setup are filled with distilled water and emptied using the ink pumps via the running background flow. This avoids ink leaking out when disconnecting the tubes from the pumps. Once all tubes no longer show any visible ink residues, the remaining water from the water reservoir may be poured into the waste container, and then all tubes can be disconnected from the connectors and pumps. The residual water remaining in the disconnected tube sections can be removed using an air-filled syringe. Finally, all used syringes, tubes, and containers are laid out to dry.

        \section{Experimental Procedure}
            \section{Measurement of the Background Flow}
To measure the background flow, please open the file Pumpensteuerung.ino and ensure that you enter a 1 in the global variable '' Versuch''.  
Now take an empty 3 ml syringe, close the opening at the bottom with your finger, and insert the tube that leads to the waste container into the syringe.  
Then enter a \(1\) in the Serial Monitor of the Arduino IDE to start the background flow, measure the time needed until \mbox{3 ml} of the syringe are filled with water, and stop the background flow with a \(0\).  
Repeat this measurement three times and enter your results into the lab protocol.  
Now calculate the average background flow rate \(Q_0\) and determine what type of flow is present.  
Additionally, determine the diffusion coefficient. Explain whether the diffusion of the ink in the water can be neglected and justify your answer.


\section{Detection of the Colors}
As already explained in the script, zeroforcing is used in this experiment. To apply it, the channel impulse response must be measured.  
For this purpose, the global variable '' Versuch''\ in the file Photosensor.ino must be set to \(1\).  
First, observe how the perceived color changes in the channel with different ink dosages and how the inks behave. To do this, you may use any injection durations between \(0-100\) ms for the different pumps.\newline
Now determine the absorption values for injection durations of 15 ms, 30 ms, 45 ms, 60 ms, and 100 ms.  
From the values for the 30 ms injection duration, construct the channel matrix \(H\), compute its pseudoinverse using the program Pseudoinverse.py, and enter this into the program for the photosensor.  
The channel matrix must be arranged in the order (yellow, magenta, cyan).  
Now change the global variable '' Versuch''\ in the photosensor control to 2.  
Inject inks with an injection duration of 30 ms into the channel again.  
What result do you expect in the ideal case, and why might your real result deviate?\newline
Also try injecting multiple colors into the channel simultaneously.  
Which threshold value should be chosen for detecting the individual colors so they remain reliably detectable?  
Which threshold should be chosen for the start flag to ensure that a 50 ms injection of yellow ink is detectable?

\section{Data Transmission}
Now that the channel matrix and all threshold values have been determined, message transmission is fundamentally possible.  
Choose arbitrary values for the waiting time after the start flag and for the bit duration at both the transmitter and receiver.  
Note that for the start flag, \(t_{Empf\text{ä}nger}= t_{Sender}-\frac{t_{Bitdauer}}{2}\) must hold, and that the bit duration should be the same for both.  
Gradually reduce the bit duration, the start flag duration, and the pump durations.  
To determine suitable threshold values, proceed as in Experiment 2.  
To send messages, change the value of Versuch from 1 to 3 in the pump control program, and from 2 to 3 in the photosensor control.  
What maximum bitrate can you achieve when you reduce the timings, injection durations, and corresponding threshold values?

\newpage
\section{Lab Protocol}
\subsection{Measurement of the Background Flow}

\vspace{1em}

\begin{table}[htbp]
\centering
\begin{tabular}{|l|l|}
\hline
\textbf{Measurement} & \textbf{Time for Background Flow Volume (3 ml)} \\
\hline
       1   &        \\
\hline
       2      &       \\
\hline
       3  &  \\
\hline
\end{tabular}
\caption{Measurement of the flow rate}
\label{apxtab:messwerte}
\end{table}

\vspace{1.5em}
\begin{tabular}{@{}l l@{}}
\textbf{Background flow:} & \underline{\hspace{2cm}} \\[1.5em]
\textbf{Average flow velocity:} & \underline{\hspace{2cm}} \\[1.5em]
\textbf{Reynolds number:} & 
\underline{\hspace{2cm}} \\[1.5em]
\textbf{Péclet number:} & \underline{\hspace{2cm}} \\
\end{tabular}
\vspace{1.5em}

\noindent
\begin{minipage}{\textwidth}
\textbf{Is the flow laminar or turbulent, and must diffusion be considered?}\\[0.5em]
\rule{\textwidth}{0.4pt}
\end{minipage}

\subsection{Detection of the Color}
\begin{table}[H]
\centering
\begin{tabular}{|c|c|c|c|c|c|}
\hline
\textbf{Time} & \textbf{15 ms} & \textbf{30 ms} & \textbf{45 ms} & \textbf{60 ms} & \textbf{100 ms} \\
\hline
 415 nm& & & & & \\ 
 \hline
 445 nm& & & & & \\ 
\hline
 480 nm& & & & & \\ 
\hline
 515 nm& & & & & \\ 
\hline
 555 nm& & & & & \\ 
\hline
 590 nm& & & & & \\ 
\hline
 630 nm& & & & & \\ 
\hline
 680 nm& & & & & \\ 
\hline
\end{tabular}
\caption{Detection of the color}
\label{apxtab:sechs_spaltend}
\end{table}

\begin{table}[H]
\centering
\begin{tabular}{|c|c|c|c|c|c|}
\hline
\textbf{Time} & \textbf{15 ms} & \textbf{30 ms} & \textbf{45 ms} & \textbf{60 ms} & \textbf{100 ms} \\
\hline
 415 nm& & & & & \\ 
 \hline
 445 nm& & & & & \\ 
\hline
 480 nm& & & & & \\ 
\hline
 515 nm& & & & & \\ 
\hline
 555 nm& & & & & \\ 
\hline
 590 nm& & & & & \\ 
\hline
 630 nm& & & & & \\ 
\hline
 680 nm& & & & & \\ 
\hline
\end{tabular}
\caption{Absorption values of Magenta}
\label{apxtab:sechs_spaltena}
\end{table}

\begin{table}[H]
\centering
\begin{tabular}{|c|c|c|c|c|c|}
\hline
\textbf{Time} & \textbf{15 ms} & \textbf{30 ms} & \textbf{45 ms} & \textbf{60 ms} & \textbf{100 ms} \\
\hline
 415 nm& & & & & \\ 
 \hline
 445 nm& & & & & \\ 
\hline
 480 nm& & & & & \\ 
\hline
 515 nm& & & & & \\ 
\hline
 555 nm& & & & & \\ 
\hline
 590 nm& & & & & \\ 
\hline
 630 nm& & & & & \\ 
\hline
 680 nm& & & & & \\ 
\hline
\end{tabular}
\caption{Absorption values of Cyan}
\label{apxtab:sechs_spaltenb}
\end{table}

\noindent
\begin{minipage}{\textwidth}
\textbf{Why is the graph for blue broader than for yellow?}\\[0.5em]
\rule{\textwidth}{0.4pt}
\end{minipage}

\vspace{2em}

\noindent
\begin{minipage}{\textwidth}
\textbf{What value should an injection of 30\,ms of ink reach after calculating the pseudoinverse, and why cannot this be measured exactly?}\\[0.5em]
\rule{\textwidth}{0.4pt}
\end{minipage}

\vspace{2em}

\noindent
\begin{minipage}{\textwidth}
\textbf{Which value should be chosen as the threshold for the colors and for the start flag?}\\[0.5em]
\rule{\textwidth}{0.4pt}
\end{minipage}

\vspace{2em}

\noindent
\begin{minipage}{\textwidth}
\textbf{Why do the values not increase linearly?}\\[0.5em]
\rule{\textwidth}{0.4pt}
\end{minipage}

\subsection{Data Transmission}

\begin{table}[H]
\centering
\footnotesize 
\begin{tabular}{|p{2cm}|p{2cm}|p{1.1cm}|p{1.8cm}|p{1.8cm}|p{1.8cm}|p{1.8cm}|p{1.2cm}|}
\hline
\textbf{Injection duration – Ink} & \textbf{Injection duration – Flag} & \textbf{Bit duration} & \textbf{Start sequence – Sender} & \textbf{Start sequence – Receiver} & \textbf{Threshold – Bit} & \textbf{Threshold – Flag} & \textbf{Data rate} \\
\hline
 & & & & & & & \\ \hline
 & & & & & & & \\ \hline
 & & & & & & & \\ \hline
 & & & & & & & \\ \hline
 & & & & & & & \\ \hline
 & & & & & & & \\ \hline
 & & & & & & & \\ \hline
 & & & & & & & \\ \hline
\end{tabular}

\caption{Data rates}
\label{apxtab:sechs_spaltenc}
\end{table}

\underline{\textbf{Best data rate achieved at:}}

\vspace{1.5em}
\begin{tabular}{@{}l l@{}}
\textbf{Injection duration – Ink:} & \underline{\hspace{2cm}} \\[1.5em]
\textbf{Injection duration – Flag:} & \underline{\hspace{2cm}} \\[1.5em]
\textbf{Sequence duration:} & \underline{\hspace{2cm}} \\[1.5em]
\textbf{Start sequence – Sender:} & \underline{\hspace{2cm}} \\[1.5em]
\textbf{Start sequence – Receiver:} & \underline{\hspace{2cm}} \\[1.5em]
\textbf{Threshold – Bit:} & \underline{\hspace{2cm}} \\[1.5em]
\textbf{Threshold – Flag:} & \underline{\hspace{2cm}} \\[1.5em]
\textbf{Data rate:} & \underline{\hspace{2cm}} \\[1.5em]
\end{tabular}

\vspace{1.5em}

\noindent
\begin{minipage}{\textwidth}
\textbf{For what reason can the data rate not be increased arbitrarily?}\\[0.5em]
\rule{\textwidth}{0.4pt}
\end{minipage}

\vspace{2em}

\noindent
\begin{minipage}{\textwidth}
\textbf{What are potential ways to further increase the data rate you achieved?}\\[0.5em]
\rule{\textwidth}{0.4pt}
\end{minipage}

        \section{Short Questions}
            \bigskip

The following short questions are to be worked on at home before carrying out the laboratory experiment and will be discussed in the colloquium beforehand.
\begin{enumerate}
    \item Determine the flow type and the diffusion coefficient for a background flow of \(Q_0 = 10 \frac{ml}{min}\) and a channel diameter of \(d_c = 2 mm\).
    \item Explain the basic principle of how zeroforcing works.
\end{enumerate}

\end{document}